\documentclass{emulateapj}
\usepackage{CJK}
\usepackage{natbib}
\usepackage{apjfonts}
\usepackage[dvips]{color}
\usepackage{graphicx}
\newcommand{\ltmark}[1]{} 
\newcommand{\Oline}[1]{\textrm{[O{\footnotesize{#1}}]}} 
\newcommand{\OlineSub}[1]{\textrm{\footnotesize[O{\scriptsize#1}]}}
\newcommand{\Hline}[1]{H{\footnotesize{#1}}} 

\newcommand{\subscript}[1]{\ensuremath{_{\textrm{\scriptsize{#1}}}}}

\def\Chandra{{\it Chandra}}
\def\Newton{{\it XMM-Newton}}
\def\XMM{{\it XMM}}
\def\Suzaku{{\it Suzaku}}
\def\Spitzer{{\it Spitzer}}
\def\Swift{{\it Swift}}
\def\INTEGRAL{{\it INTEGRAL}}
\def\BeppoSAX{{\it BeppoSAX}}
\def\FeKa{Fe K$\alpha$}

\shorttitle{Outflowing Corona}
\shortauthors{Liu et al.}

\begin{document}
\begin{CJK*}{UTF8}{gkai}

\title{Are X-ray Emitting Coronae around Super Massive Black Holes Outflowing?}
 
\author{Teng Liu (刘腾), Jun-Xian Wang (王俊贤), Huan Yang (杨欢), Fei-Fan Zhu (朱飞凡), You-Yuan Zhou (周又元)}
\affil{CAS Key Laboratory for Research in Galaxies and Cosmology, Department of Astronomy, University of Science and Technology of China, Hefei, Anhui 230026, China; liuteng@ustc.edu.cn, jxw@ustc.edu.cn}

\begin{abstract}
Hard X-ray emission in  radio-quiet active galactic nuclei (AGNs) is believed  to be produced via inverse Compton scattering by hot and compact coronae near the  super massive black  hole.
However the origin and physical properties of the coronae, including geometry, kinematics and dynamics, yet remain poorly known. 
In this work, taking \Oline{IV} 25.89 $\micron$ emission line as an isotropic indicator of AGN's intrinsic luminosity, 
we compare the intrinsic corona X-ray emission between Seyfert 1 and Compton-thin Seyfert 2 galaxies,  which are viewed at different inclinations according to the unification scheme. 
We compile a sample of 130\ltmark{84Sy1+46Sy2 including OIV upper limits, 80Sy1+45Sy2 excluding OIV upper limits} Compton-thin Seyfert galaxies with both \Oline{IV} 25.89 $\micron$ line luminosities measured with \Spitzer-IRS and X-ray spectra observed by \Newton{}, \Chandra{}, \Suzaku{} or \Swift{}.
Known radio-loud sources are excluded.
We fit the X-ray spectra to obtain the absorption-corrected 2 -- 10 keV continuum luminosities.
We find that Seyfert 1 galaxies are intrinsically brighter in intrinsic 2 -- 10 keV emission by a factor of $2.8^{+0.5}_{-0.4}$
($2.2^{+0.9}_{-0.3}$ in \Swift-BAT 14 -- 195 keV emission), comparing with Compton-thin Seyfert 2 galaxies. 
The Seyfert 1 and Compton-thin Seyfert 2 galaxies follow a statistically identical correlation between the absorption-corrected 2 -- 10 keV luminosity and the 14 -- 195 keV luminosity, indicating that our absorption correction to  the 2 -- 10 keV  flux is sufficient.
The difference in X-ray emission between the two populations is thus unlikely due to X-ray absorption, and instead  implies an intrinsic anisotropy in the corona X-ray emission.
 This striking anisotropy of X-ray emission can be explained by a bipolar outflowing corona with a bulk velocity of $\sim 0.3 - 0.5 c$.
 This would provide a natural link between the so-called coronae and weak jets in these systems.
Other consequences of  outflowing coronae are also discussed.
\end{abstract}

\keywords{ galaxies: active -- galaxies: nuclei -- galaxies: Seyfert -- galaxies: jets -- X-rays: galaxies -- relativistic processes }

\section{INTRODUCTION}
Strong hard X-ray emission has been detected in black hole accreting systems spanning from stellar mass ($\sim 10 M_{\sun}$) to super massive scales ($\sim 10^5 - 10^{10} M_\sun$).
Such systems are  the  dominant population in  the X-ray sky.
In radio-quiet super massive black hole (SMBH) accretion systems, the primary hard X-ray emission is power-law shaped in  spectrum with a high energy cutoff at $\gtrsim 100$ keV \citep{Molina13,Ricci11}.
It is now widely accepted that X-ray emission in radio-quiet AGNs comes from inverse Compton scattering of the seed photons from the accretion  disks through hot and compact  plasma, named corona \citep{Haardt91,Haardt93}.
Such coronae are also required in radio-loud AGNs, in which strong relativistic jets could produce extra emission from radio to X-ray and even gamma-ray, and in stellar mass black hole accretion systems \citep[e.g.][]{Plotkin12}.
However, we still know very poor about the origin and physical nature of the corona.

Being free from dust extinction, hard X-ray emission has been widely used to represent the intrinsic power  of AGNs, e.g., to measure the ratio of type 1 to type 2 AGNs and to derive the  AGN bolometric luminosity \citep{Melendez08,Rigby09,Burlon11,Malizia09,Runnoe12}.
However, we still know little observationally to what level the corona X-ray emission is isotropic.
While a slab-like corona could produce weaker X-ray emission at higher inclination due to  projection effect \citep[e.g.][]{Zhang05,Chen13}, the suggested patchy structure and small opacity of the corona indicate the anisotropy produced this way is rather weak \citep{Haardt94,Zdziarski00,Molina09}.
Meanwhile, similar to solar  coronae, the X-ray coronae in black hole systems might be heated through magnetic processes and could produce magnetic inflow/outflow above the disk \citep{Beloborodov99}.
Bulk motion of the  flow, if fast enough, would produce beamed X-ray emission.
The X-ray emission is also likely produced in the jet base, which is physically similar to an outflowing corona \citep{Markoff05}.
Light bending effect of the black hole could also produce anisotropy in X-ray emission, by reducing the X-ray flux escaping upward \citep{Miniutti04,Chen13}.  
These arguments  suggest that the corona X-ray emission in AGNs is likely anisotropic, although weak if there is any as no observational evidence has yet been reported.
However, measuring the level of the anisotropy or putting strong constraint to it is essential to  understand the origin and physical properties of the corona.  

In the unification scheme of AGNs, type 1 sources are viewed face-on while type 2  sources viewed edge-on \citep{Antonucci93}, providing an opportunity to  measure the anisotropy of the corona X-ray emission.
In this work we present an observational investigation to compare the corona X-ray emission in type 1 and type 2 AGNs at given intrinsic black hole accretion power. 
To do so, we need an independent AGN luminosity indicator and samples of AGNs with different inclination angles.

Inspired by recent studies \citep{Melendez08,Rigby09,Diamond09,Weaver10,LaMassa10,Liu10}, we opt to use \Oline{IV} $25.89 \micron$ line emission, a forbidden line produced in the so called narrow line region (NLR),  as an intrinsic luminosity proxy of AGNs.
Comparing with the widely used \Oline{III} $5007$\AA\ line, \Oline{IV} line is significantly less attenuated by dust extinction (A$_V$ $\sim$ 39 corresponds to A\subscript{25.89 \micron} $\sim$ 0.06 - 0.18; Goulding \& Alexander 2009), and has relatively higher ionization potential (54.9 eV) thus is less affected by contamination from star formation in the host galaxy.   
The validity of using \Oline{IV}\ as intrinsic AGN luminosity indicator has been confirmed observationally \citep{Diamond09,Melendez08,Rigby09}.  Particularly, Diamond-Stanic et al. (2009) showed that the \Oline{IV}\ luminosity distributions are indistinguishable for obscured and unobscured AGNs in a well-defined  galaxy-magnitude-limited sample, while \Oline{III} luminosities are systematically lower in obscured sources.
However we note that unlike \Oline{III}, measurements of \Oline{IV}\ line fluxes are available in much fewer AGNs (see \S \ref{reduction}).

We select to compare corona X-ray emission of type 1 AGNs with that of Compton-thin type 2 AGNs.
Only securely identified Compton-thin sources (with X-ray obscuration column density $N_H < 10^{24}\ cm^{-2}$) are included, because for Compton-thick ones reliable correction to X-ray obscuration is extremely hard or even impossible.
Known radio-loud sources are also excluded to avoid contaminations from known strong jet emission. 
 2 -- 10 keV intrinsic (i.e., absorption-corrected continuum) luminosities are obtained through spectral fitting. 

In \S \ref{reduction}, we describe the samples, and the processing of X-ray data.
In \S \ref{statistic}, we compare the intrinsic X-ray emission between type 1 and type 2 subsamples at given \Oline{IV} luminosity, showing relatively weaker intrinsic X-ray emission in type 2 sources.
In \S \ref{discuss}, we present discussions on the robustness of  the anisotropy of corona X-ray emission, and finally interpret it in terms of corona outflowing.
\S \ref{sec:summary} presents the summary.

\section{SAMPLE AND DATA REDUCTION}
\label{reduction}
\subsection{\Oline{IV} Sample Compilation\label{o4sample}}
\label{O4sample}
To gather a large sample of AGNs with both \Oline{IV}  flux measurements and X-ray observations, we first combine five major samples  in literature with \Oline{IV} $25.89 \micron$ line  fluxes measured by \Spitzer{} Infrared Spectrometer \citep[IRS,][]{Diamond09,Weaver10,Tommasin10,Dasyra11,Pereira10}.
The five samples are briefly described as follows:
\begin{description}
\item[A]
A subsample of 91 Seyfert galaxies with \Spitzer-IRS high-resolution spectroscopic observations \citep{Tommasin10} from the $12\micron$ flux limited sample of Seyfert galaxies \citep[116 sources,][]{Rush93}.
\item[B]
\Spitzer-IRS low-resolution spectroscopic observations of a spectroscopically selected, galaxy-magnitude-limited sample of Seyfert galaxies from the revised Shapley-Ames (RSA) catalog \citep{Diamond09}.
The RSA Seyfert sample, containing 18 type 1 and 71 type 2 Seyfert galaxies, is a well studied galaxy-magnitude-limited complete sample \citep{Maiolino95,Ho97}.
\item[C]
A subsample of 79 AGNs with high-resolution \Spitzer-IRS spectroscopy from the \Swift-BAT hard X-ray selected  local AGN sample \citep{Weaver10}. 
\item[D]
\Spitzer-IRS high-resolution spectroscopy observations of a composite AGN sample \citep{Dasyra11}.  A list of optical spectroscopic catalogs of AGNs with existing M\subscript{BH} estimates and \Oline{III} $5007$\AA\ detections are involved in matching with all the IRS targets.
The final sample consists of 81 AGNs with resolved [S{\footnotesize IV}], [Ne{\footnotesize III}], \Oline{IV} or [Ne{\footnotesize V}] lines. 
\item[E]
A collection (from literature, archive, and the author's own observing programs) of \Spitzer-IRS high-resolution spectra of 426 galaxies, including quasars, Seyferts, LINERs and \Hline{II} galaxies \citep{Pereira10}.  
\end{description}

These samples somehow overlap with each other.
High-resolution data are adopted with higher priority when available. 
High-resolution \Oline{IV} flux measurements from different subsamples for overlapped sources are generally consistent with each other.
In a few of them (7 sources, 5\% of the final sample), the differences are as large as by a factor of $>$1.5. 
We simply adopt the average \Oline{IV} fluxes in logarithm space for these overlapped sources.
The \Oline{IV} fluxes  are converted into luminosities adopting the same cosmological parameters\footnote{$H_0 = 70\ km\ s^{-1}\ Mpc^{-1}$, $\Omega_m = 0.27$, and $\Omega_\lambda = 0.73$, adopted throughout this paper.}.

We then match the  composite \Oline{IV} sample to \Newton{} data archive to search for reliable X-ray spectra.
For a few sources without \XMM{} observations, we supplement  the X-ray spectra collection with \Chandra{}, \Suzaku{} or \Swift{} data (see Table \ref{tab:data}).
We perform literature search and independent X-ray spectral fittings  to get the classifications of their X-ray obscuration nature.
Possible Compton-thick sources are excluded, and only securely identified Compton-thin ones are included.
The criteria we adopted to identify secure Compton-thin sources are:
1) X-ray spectra fitting could provide reliable measurements to X-ray  obscurations at $N_H < 10^{24}\ cm^{-2}$, if with high quality spectra;
2) for those sources without  good-enough X-ray spectra, sources with T ratio $< 1$ (T ratio = $f_{2-10 keV}/f_\OlineSub{III}$, where $f_{2-10 keV}$ is the observed X-ray flux, and $f_\OlineSub{III}$ the dust extinction corrected line flux) or narrow \FeKa{} line EW $> 600$ eV are excluded as candidates of Compton-thick sources \citep{Bassani99}.
This is a rather conservative approach since some known Compton-thin sources also show T ratio $< 1$ or line EW $> 600$ eV \citep{Bassani99}.
We exclude 19 sources (15 of which are type 2) with too few X-ray counts ($< 100$ in 2 -- 10 keV) in \XMM{} data, for which reliable measurements to the X-ray  obscurations are hard.
To avoid contamination from strong relativistic jet to the X-ray emission, known radio-loud sources are excluded.
In Table \ref{tab:data} we present the sample with references to their X-ray identifications.
The full sample including possible Compton-thick sources will be presented in a future paper.  

We also excluded 5 X-ray un-obscured Seyfert 2 galaxies, i.e. those with X-ray obscuration column  densities $N_H < 10^{22}\ cm^{-2}$ but without any evidence of broad line regions (BLRs) detected, which could be physically different from normal Seyfert 2 galaxies \citep{Brightman08}.
Including them into analyses would however slightly strengthen our conclusion in this work. 

Finally, the  composite sample includes 84 type 1 AGNs (including Seyfert 1, Seyfert 1.2 and Seyfert 1.5 galaxies, hereafter Sy1s) and 46 Compton-thin type 2 AGNs (including Seyfert 1.8, Seyfert 1.9 and Seyfert 2, hereafter Sy2s).

\subsection{X-ray Data Reduction and Spectral Fitting}
To obtain the intrinsic 2 -- 10 keV luminosities of sources in our sample, we first  collect data from literature (see Table \ref{tab:data} and Appendix \ref{Apd:collection}).  
For a major fraction of the sample we need to perform our own spectral fitting to measure the intrinsic X-ray fluxes as either the X-ray data  are not published yet, or no intrinsic 2 -- 10 keV fluxes are given in literature.
For sources with multiple \XMM{}  observations, only the observations during which the sources were Compton-thin are kept, and exposure-time weighted mean intrinsic luminosities are  calculated in logarithm space.

The \Newton{} data are processed with the \XMM-SAS package in the standard way.
The source regions to extract the spectra  are optimized with the ``eregionanalyse'' task.
Background extraction regions are a few manually selected circles with proper radii around the source region, being central symmetric as much as possible, and kept away from the CCD edges, the out-of-time events strips and other sources.
Each exposure is checked for pile-up with the task ``epatplot''.
A core circle region with a small radius is excluded in the spectra extracting in case of pile-up.
The circle radius is manually selected by increasing step by step until the pile-up effect no longer shows up in the ``epatplot'' result.
We generate the source and background spectra together with the appropriate redistribution matrix and ancillary response file from the source and background regions for each exposure, using the task ``especget''.
The \Chandra{} and \Swift-XRT data of a few sources  are also processed with their standard pipelines.

 To measure the absorption-corrected power-law flux in the 2 -- 10 keV band, we fit the 0.5 -- 10 keV spectra in ``Xspec'' with a power-law absorbed by both Galactic and intrinsic absorbers, together with three additional emission components --  a soft excess,  a cold reflection component and a narrow \FeKa{} line at 6.4 keV.
We are only interested in decomposing the additional components from the power-law, but not in the complex origins of soft excess and reflection.
For Sy1s, we fit the soft excess with the traditionally used blackbody model, while for Sy2s with an extra soft power-law model, because the soft excess displays different origins for unabsorbed and absorbed AGNs \citep{Corral11}.
In some cases, additional blackbody or power-law or  ionized gas emission is needed to model the soft excess.
Because the spectral qualities of most sources are not good enough to put reasonable constrains on the parameters of the reflection component (``pexrav'' in ``Xspec''), we  fix the cutoff energy of the power-law at 200 keV, and the reflection fraction $R$ at $0.5$.
Adopting different values of $R$ does not alter the results in this work (see \S \ref{absorptioncorrection}).
In a few cases, a broad \FeKa{} line is also needed to improve the fitting statistics.

\section{COMPARISON BETWEEN TYPE 1 AND TYPE 2 AGNS}
\label{statistic}
\begin{figure}
\centering
\plotone{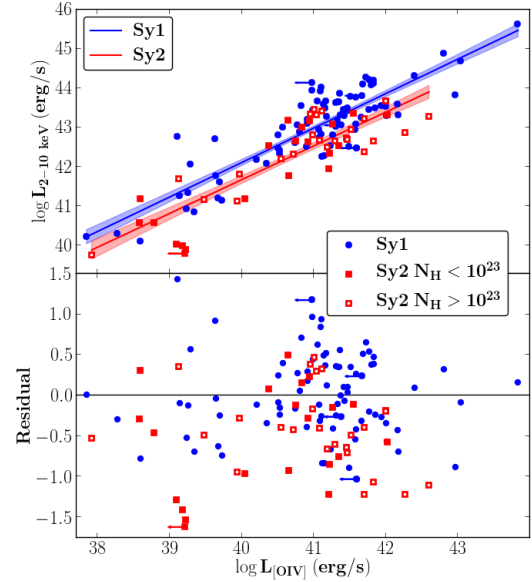}
\caption{The  relations between  the 2 -- 10 keV intrinsic X-ray emission (absorption-corrected power-law emission) and the \Oline{IV} $25.89 \micron$ line emission for Sy1s (type 1 Seyfert galaxies, blue dots) and Compton-thin Sy2s (type 2 Seyfert galaxies, red squares).
Sy2s are divided into two types with $N_H$ lower (solid squares) or higher (open squares) than $10^{23}\ cm^{-2}$.
A couple of upper limits to \Oline{IV}  luminosities are plotted in arrows.
Lines plot the best-fit correlations  (through simple linear regression), and shaded regions  are the $1\sigma$ confidence bands of the fits.
The typical (median) errors in $L_{\OlineSub{IV}}$ and $L_{2-10keV}$ are 5\% and 4\% respectively, and are not plotted for simplicity.
In the lower panel we plot the residuals of the data points to the best-fit line of Sy1s (along the y-axis) for both Sy1s and Sy2s.
Sy2s are obviously weaker in intrinsic 2 -- 10 keV emission.
\label{2-10-OIV}}
\end{figure}

In Fig. \ref{2-10-OIV}, we plot the intrinsic 2 -- 10 keV X-ray luminosities versus \Oline{IV} luminosities of our sample.
Clear correlations between intrinsic 2 -- 10 keV X-ray and \Oline{IV} line emission, both of which somehow reflect the intrinsic power of the SMBH accretion, are detected for both type 1 and type 2 AGNs, confirming previous studies \citep{Melendez08,Rigby09,Diamond09,Weaver10}.
Simple linear regression \citep[][taking \Oline{IV} as the independent variable]{Isobe90} is performed on the Sy1s and Sy2s subsamples.
The best-fitted lines are:
\[
Sy1s:	y = (0.88\pm0.05) \times (x-41.17) + 43.13\pm0.06
\]
\[
Sy2s:	y = (0.86\pm0.08) \times (x-40.99) + 42.52\pm0.09
\]
Remarkably weaker intrinsic X-ray emission at given \Oline{IV} luminosity in type 2 AGNs is seen in this figure.
We perform K-S test to examine whether the residuals of data points from Sy1s and Sy2s to the best-fit slope of Sy1s (lower panel of Fig. \ref{2-10-OIV}) are extracted from the same population.
At a confidence level of 99.94\% they are not.
The deviation of residuals of the two subsamples shows that Compton-thin type 2 AGNs are fainter in intrinsic  2 -- 10 keV emission by a factor of $2.8^{+0.5}_{-0.4}$, comparing with Sy1s.
We note  that orthogonal distance regression  \citep[ODR,][]{Isobe90} yields different correlation slopes, but does not alter other results in this work.
The best-fitted ODR lines are:
\[
Sy1s:	y = (1.00\pm0.06) \times (x-41.17) + 43.15\pm0.06
\]
\[
Sy2s:	y = (0.99\pm0.08) \times (x-40.99) + 42.56\pm0.09
\]

\begin{figure}
\centering
\plotone{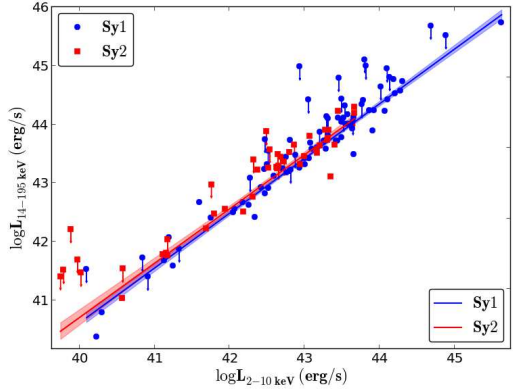}
\caption{\Swift-BAT 14 -- 195 keV versus intrinsic 2 -- 10 keV luminosities.
Sy1s and Compton-thin Sy2s follow a statistically identical tight correlation.
Arrows plot the upper limits to BAT non-detected sources.
Lines plot the best-fit correlations, and shaded regions the $1\sigma$ confidence bands of the fits.
\label{14-195-2-10}}
\end{figure}

In our statistical analyses, we omit 5 sources (4 Sy1s and 1 Sy2, plotted in Fig. 1) with only \Oline{IV} upper limits.
These \Oline{IV} upper limits could be taken into account with survival analyses by either taking the intrinsic 2 -- 10 keV luminosity as the independent variable, or treating the upper limits to \Oline{IV} luminosities as lower limits to 2 -- 10 keV emission instead.
Both approaches yield consistent results.
Here we adopt the more straightforward approach by excluding the \Oline{IV} upper limits and taking \Oline{IV} as the independent variable.

Hard X-ray photons above 10 keV are insensitive to Compton-thin obscuration.
We match our sample to \Swift-BAT 70 month catalog \citep{Baumgartner13} to extract their 14 -- 195 keV X-ray fluxes.
For \Swift-BAT non-detected sources, we adopt an upper limit of $1.34 \times 10^{-11}\ erg\ cm^{-2}\ s^{-1}$ to their 14 -- 195 keV fluxes, which is the $5\sigma$ sensitivity limit of \Swift-BAT all sky survey for 90\% of the sky \citep{Baumgartner13}. 
We plot 14 -- 195 keV versus intrinsic 2 -- 10 keV luminosities in Fig. \ref{14-195-2-10}, and perform Buckley-James linear regression on type 1 and type 2 subsamples, taking account of the upper limits.
Both type 1 and type 2 sources follow a statistically identical tight correlation.
The best-fitted lines in Fig. \ref{14-195-2-10} are:
\[Sy1s: y = (0.93\pm0.03) \times (x-43.23) + 43.63\pm0.03\]
\[Sy2s: y = (0.92\pm0.05) \times (x-42.63) + 43.14\pm0.04\]
This confirms that both quantities measure the intrinsic corona emission.
It also proves that our absorption corrections to  the 2 -- 10 keV  fluxes are sufficient, otherwise we would see weaker 2 -- 10 keV emission (relative to 14 -- 195 keV emission) in Sy2s.

\begin{figure}
\centering
\plotone{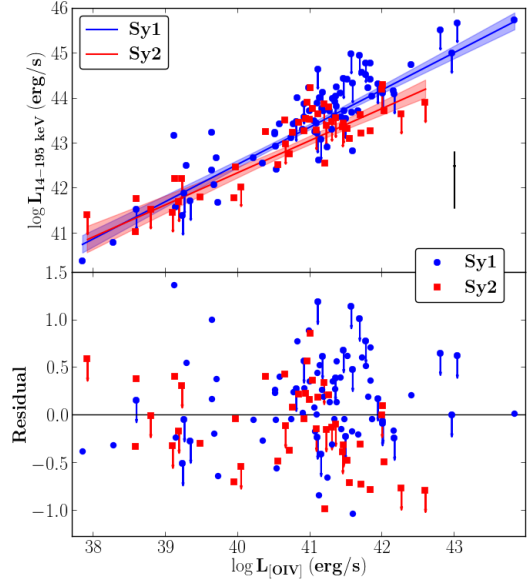}
\caption{Similar to Fig. 1, but  replacing the intrinsic 2 -- 10 keV X-ray luminosity with  the \Swift-BAT 14 -- 195 keV luminosity.
The typical error in $L_{14-195keV}$ is plotted at the lower right corner in the upper panel.
Arrows plot the upper limits to BAT non-detected sources.
\label{14-195-OIV}}
\end{figure}

Comparing 14 -- 195 keV emission with \Oline{IV} emission (Fig. \ref{14-195-OIV}) well confirms the pattern shown in Fig. \ref{2-10-OIV}.
Residuals  for Sy1s and Sy2s are calculated in the same way as above (lower panel in Fig. \ref{14-195-OIV}) and then compared with logrank survival analysis.
We find a significant difference at a confidence level of 99.1\%, and a flux decrement in type 2 sources by a factor of $2.2^{+0.9}_{-0.3}$.
The best-fitted (Buckley-James regression) lines in Fig. \ref{14-195-OIV} are:
\[Sy1s: y = (0.83\pm0.06) \times (x-41.17) + 43.50\pm0.06\]
\[Sy2s: y = (0.71\pm0.10) \times (x-40.99) + 43.06\pm0.09\]
Sy2s tend to have a slightly flatter correlation slope  (though statistically insignificant) comparing with Sy1s.
It is possible that the difference between Sy1s and Sy2s is more significant at high luminosities.
However much larger samples are needed to demonstrate this postulation.

\section{DISCUSSION}
\label{discuss}
\subsection{Hints in Previous Studies}
It is widely assumed that the corona emission in AGNs is isotropic.
However, low to moderate level of anisotropy can't be ruled out based on previous studies.
Actually, in several studies hints for anisotropic X-ray emission are visible. 

\citet{Rigby09} compared the \Oline{IV} $25.89 \micron$ emission with  the 10 -- 200 keV X-ray emission for  the RSA Seyfert sample.
They found that Compton-thin Sy2s are weaker in 10 -- 200 keV emission comparing with Sy1s by a factor of $1.9 \pm 0.5$.
This decrement in 10 -- 200 keV emission in Compton-thin Sy2s,   although statistically insignificant (91\%), is in good agreement with ours.
However it is hard to attribute it to obscuration since it would require an average obscuration of $\log N_H = 24.3\pm0.1\ cm^{-2}$, which is too high comparing with the median column density for their Compton-thin sample ($\log N_H = 23.0\ cm^{-2}$) measured based on 2 -- 10 keV spectral fitting.
We note that the smaller sample size of \citet{Rigby09} and the fact that a significant fraction of their sources only have upper limits to 10 -- 200 keV emission might have prevented them to detect a statistically significant difference between Sy1s and Compton-thin Sy2s.
A similar trend was also seen in \citet{Liu10}.
By utilizing \Oline{IV} emission as intrinsic luminosity indicator, they found that the monochromatic 6.4 keV continuum in Sy1s is $4.2 \pm 1.6$ times stronger than Compton-thin Sy2s, which on average  requires a $N_H = 8.6\pm1.9 \times 10^{23}\ cm^{-2}$ in Compton-thin Sy2s, similar to the  value reported by \citet{Rigby09}.

\subsection{On the Isotropy of \Oline{IV} $25.89 \micron$ Line Emission}

Although \Oline{IV} $25.89 \micron$ emission line is believed to be a good isotropic indicator of AGN intrinsic luminosity \citep{Diamond09,Melendez08,Rigby09}, below we discuss possible factors which might affect its isotropy.

\citet{Zhang08} found that Sy1s have smaller [N II]/H$\alpha$ ratios on the BPT diagram comparing with Sy2s.
They proposed that the inner  NLR of AGN is likely heavily obscured in type 2 sources by the extending region of the torus.
This effect, if confirmed, could yield weaker \Oline{IV} $25.89 \micron$ and \Oline{III} emission in Sy2s due to heavy  obscurations. 

Another factor is that,  AGN with smaller covering  factor of the obscuring torus (and thus larger open-cone angle of the NLR),  is more likely to be viewed as Type 1, than  that with larger torus covering factor \citep{Turner09}.
Naturally, Sy1s tend to have stronger NLR emission than Sy2s \citep{Lawrence13}.

\begin{figure}
\plotone{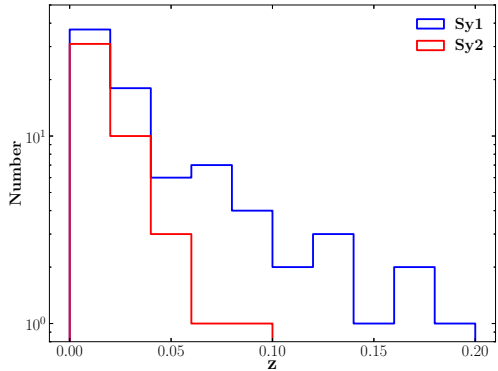}
\caption{Redshift distributions of Sy1s and Compton-thin Sy2s.
\label{redshift}}
\end{figure}

Possible slit loss in \Oline{IV} measurements might also lead observational bias if Sy1s and Sy2s have different redshift distributions.
In Fig. \ref{redshift} we see that Sy2s tend to have lower redshifts in our  composite sample, thus slit loss could be more severe in Sy2s than in Sy1s. 
But note slit loss in \Spitzer{} spectra could be rather weak anyway due to the large \Spitzer-IRS slit width (4.7 -- 11.1\arcsec\ for observations adopted in this work).
Furthermore, the \Oline{IV} emission region was found to be more compact that \Oline{III} \citep{Dasyra11}.

However, all three possible factors above lead to higher \Oline{IV} emission in Sy1s comparing with Sy2s, therefore could not ease the relative stronger intrinsic hard X-ray emission in Sy1s that we have detected.
Instead, our findings would be further strengthened.

On the contrary, could the \Oline{IV} emission be in fact intrinsically stronger in Sy2s than in Sy1s?
Based on the large dispersion in the  ratios of \Oline{III} to hard X-ray luminosities, \citet{Trouille10} postulated that some AGNs may have low \Oline{III} luminosities because of the complexity of the NLR structure.
They also suggested  that \Oline{IV} emission may be similarly affected by the NLR structure complexity.
Is it possible that the NLR of Sy2s absorb a larger fraction of ionization emission -- because of more gas in Sy2s, either around the nuclei or at pc--kpc scales in the host galaxies -- thus provide stronger narrow emission lines?
\citet{Diamond09} showed that the \Oline{IV}\ luminosity distributions are indistinguishable for obscured and unobscured AGNs  in a complete AGN sample -- the RSA Seyfert sample.
This provides  a strong support to the isotropy of \Oline{IV} emission in the sense that Sy1s and Sy2s have intrinsically similar \Oline{IV} emission.
Below we extend such test to examine whether weak anisotropy in \Oline{IV} emission could be tolerated by current data.
Considering the best-fitted slope of Sy1s in the intrinsic 2 -- 10 keV luminosity -- \Oline{IV} luminosity distribution (0.88, see \S \ref{statistic}), a 2.8 times decrement in the X-ray emission of Sy2s can be explained by a 3.2 times increment in their \Oline{IV} emission.
We test this possibility by comparing the \Oline{IV} luminosity distributions of Sy1s and Sy2s within two complete AGN samples -- the 12$\micron$ Seyfert sample \citep{Rush93,Tommasin08,Tommasin10} and the RSA Seyfert sample \citep{Maiolino95,Ho97,Diamond09}, as mentioned in \S \ref{O4sample}.
Both samples are highly complete in \Oline{IV} line measurements in our  composite \Oline{IV} sample, with 113 out of 116 sources of the 12$\micron$ sample and all 89 of the RSA sample.
We perform logrank tests on the \Oline{IV} luminosity distributions of Sy1s and Sy2s within the two samples, considering a few upper limits  to \Oline{IV} luminosities.
We find no statistical  difference in \Oline{IV} luminosity distributions  between Sy1s and Sy2s,
confirming the isotropy of \Oline{IV} line emission as reported by \citet{Diamond09}.
However, if correcting the assumed 3.2 times increment in the \Oline{IV} fluxes of Sy2s, the distributions become significantly different between Sy1s and Sy2s 
(at a confidence level of 99.6\% for  the 12$\micron$ sample and 96.6\% for  the RSA sample respectively).
We conclude that Sy2s unlikely have relatively stronger \Oline{IV} emission comparing with Sy1s.

\subsection{On the Sample Incompleteness and Bias\label{sampleincompletenessbias}}
The sample we complied is based on five \Oline{IV} subsamples, thus is not homogeneous.
We test the robustness of our results with the two complete samples used above -- the 12$\micron$ Seyfert sample and the RSA Seyfert sample,  both of which are highly complete in \Oline{IV} measurements.
Their X-ray completenesses are also high, with available X-ray data for 91 out of the 113 12$\micron$ sources and 75 out of  the 89 RSA sources.
Also, the X-ray incompleteness is more severe to type 2 AGNs as obscured sources are much fainter in X-ray, therefore correcting such incompleteness (if possible in the future) would further strengthen our finding.
For the 12$\micron$ sample only, we find Sy2s are fainter in intrinsic 2 -- 10 keV emission at given \Oline{IV} luminosity comparing with Sy1s by a factor of $2.5^{+0.7}_{-0.5}$ with a confidence level of 95.4\% (38 Sy1s versus 18 Compton-thin Sy2s).
Considering only the RSA sample, we obtain a factor of $2.2^{+0.6}_{-0.5}$ with a confidence level of 89.7\% (22 Sy1s versus 17 Compton-thin Sy2s).
Combining the 12$\micron$ sample and the RSA sample we obtain a factor of $2.4^{+0.5}_{-0.5}$ with a confidence level of 97.9\% (44 Sy1s versus 29 Compton-thin Sy2s)
Considering the  remaining sources in our  composite sample we obtain statistically consistent numbers, a factor of $3.2^{+0.9}_{-0.8}$ with a confidence level of 96.6\% (36 Sy1s versus 16 Compton-thin Sy2s).
Therefore our results are independent  of the selection of subsamples. 

Another issue is the possible bias in the optical identifications of Sy1s and Sy2s due to strong nuclei  variations in some sources.
In other words, a Sy1 could be mis-identified as Sy2 if its nuclei activity  significantly reduced recently, yielding much weaker broad emission lines and X-ray emission but relatively stronger \Oline{IV} emission, since \Oline{IV} takes  a longer time to respond to nuclei variation. 
However,  sources with such state transitions would more likely be identified as Compton-thick in X-ray, due to the significant weakness of the central X-ray emission \citep{Guainazzi2005}, which would have been excluded from our sample.
To further address this issue, we re-divide our whole sample into two subsamples with $N_H<10^{22}\ cm^{-2}$ (66 sources) and $N_H>10^{22}\ cm^{-2}$ (59 sources) respectively, independent  of their optical identifications.
Following the method as described in \S \ref{statistic}, we find that Seyfert galaxies with $10^{22}\ cm^{-2}$ $<$ $N_H$  $<$ $10^{24}\ cm^{-2}$ have weaker intrinsic X-ray emission at given \Oline{IV} luminosity than sources with lower $N_H$ (by a factor of $1.70_{-0.25}^{+0.30}$, with a confident level of 99.5\%).
This demonstrates  that the possible bias in optical identifications is unable to erase the observed difference between Sy1s and Sy2s in our sample. 

\subsection{On the Absorption Correction\label{absorptioncorrection}}
In \S \ref{statistic}, we have shown that our absorption correction to  the 2 -- 10 keV  flux is generally sufficient, as Sy1s and Sy2s follow a statistically identical correlation between intrinsic 2 -- 10 keV luminosity and 14 -- 195 keV luminosity. 
We also found that both intrinsic 2 -- 10 keV emission and 14 -- 195 keV emission are relatively weaker in Sy2s than in Sy1s at given \Oline{IV} luminosity. 
We note previous studies also reported that Sy2s have relatively weaker hard ($>$ 10 keV) X-ray emission than Sy1s \citep{Rigby09,Weaver10,LaMassa10}, and interpreted this result in terms of heavy X-ray  absorptions.
Particularly, \citet{Rigby09} found that known Compton-thin Sy2s are weaker in  14 -- 195 keV emission comparing with Sy1s by a factor of 1.9 $\pm$ 0.5, and attributed it to  a possible obscuration which is much heavier than values obtained through spectral fitting and could attenuate even 14 -- 195 keV X-ray emission. 
However, such interpretation is unlikely for Compton-thin Sy2s, otherwise peculiar thick absorbers which are  essentially impenetrable to X-rays would be required in most (if not all) Compton-thin Sy2s, and such absorbers can not be accounted for with current common X-ray absorption fitting procedures.

One may suspect that the absorption correction might be inadequate for heavily obscured Sy2s (i.e. those with $10^{23}<N_H<10^{24}\ cm^{-2}$), as their  obscurations might have been significantly underestimated.
However, in Fig. \ref{2-10-OIV} we  show that Sy2s with $10^{23}<N_H<10^{24}\ cm^{-2}$ (24 out of 45) are statistically indistinguishable from Sy2s with $N_H<10^{23}\ cm^{-2}$ in terms of their intrinsic 2 -- 10 keV emission (relative to \Oline{IV}), ruling out such possibility.

Below we present further discussions  on issues related to absorption correction.

We use the traditional photoelectric absorption model ``wabs'' in ``Xspec'' in spectra fitting.
Using a more modern obscuration model ``tbabs'' instead of ``wabs'' does not affect the absorption-corrected fluxes.
The mean ratios of the absorption-corrected fluxes using two different models are 99.97\% and 99.76\% for Sy1s and Sy2s respectively.

\begin{figure}
\centering
\plotone{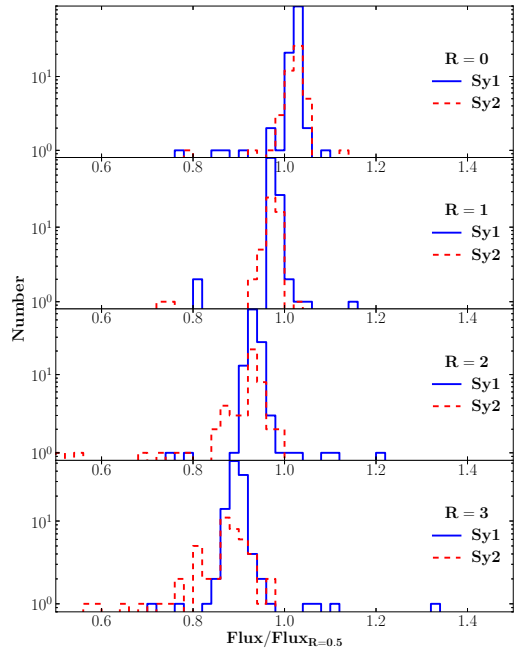}
\caption{Distributions of the ratios of the intrinsic 2 -- 10 keV fluxes  measured with $R=0,1,2,3$ to  those  measured with $R=0.5$.
\label{refl_hist}}
\end{figure}

During spectra fitting, we  fix the reflection fraction $R$ of the $pexrav$ model at 0.5.
We show in Fig. \ref{refl_hist} that adopting different $R$ values only produces minor changes to the intrinsic 2 -- 10 keV fluxes, comparing  with the difference we  find between Sy1s and Sy2s.
Also, it was suggested that reflection in Sy2s is likely stronger than in Sy1s \citep{Ricci11}, thus adopting a uniform $R$ for both Sy1s and Sy2s would lead to a systematic (although rather weak) overestimation of the intrinsic X-ray emission in Sy2s.
Correction to this effect could further strengthen  our result.

In a couple of type 1 AGNs, hard X-ray excesses were detected above 20 keV, suggesting they are partially obscured by Compton-thick clouds \citep{Turner09,Reeves09,Risaliti09}, and such partially covering absorber could be common in type 1 AGNs \citep{Tatum13}.
X-ray spectral fitting at $< 10$ keV thus could have underestimated the intrinsic 2 -- 10 keV emission in AGNs.
This effect, however does not affect the major results in this work, since the partially covering absorber, if ubiquitous, produces consistent biases to type 1 and type 2 sources, since we have shown they follow an identical correlation between 14 -- 195 keV and intrinsic 2 -- 10 keV emission (see Fig. \ref{14-195-2-10}).
The biases could have been canceled while we compare between them.
Actually, the average ratio of 14 -- 195 keV flux to intrinsic 2 -- 10 keV flux  in our sample is consistent with a power-law spectrum with a photon index $\Gamma = 1.78^{+0.18}_{-0.37}$ ($\Gamma = 1.87^{+0.17}_{-0.38}$ assuming a power-law spectrum with a reflection fraction $R=0.5$), suggesting that partial covering absorption effect is weak in our sample. 

\citet{Tombesi10} reported detections of highly ionized X-ray  obscurations in local AGNs with $N_H \sim 10^{22} - 10^{24}\ cm^{-2}$, which can't be corrected through our spectral fitting.
However, most of the ionized absorbers are not thick enough to significantly attenuate the 2 -- 10 keV continuum through Compton scattering.
Furthermore, the similar detection  rates of such absorbers in type 1 and type 2 AGNs  suggest that they have no favored inclination.
 Their large sky coverage, if confirmed with high S/N detection, also indicates  that they would not produce inclination dependent biases to X-ray fluxes.

\subsection{Outflowing Corona}
It was proposed that there could be projection effect in corona's X-ray emission, as if the corona is flat and optically thick, it will produce weaker radiation at larger inclination angle \citep{Zhang05}.
However, studies have shown that the corona can only cover a minor fraction of the disk, and should be patchy instead of flat/disk-like \citep{Haardt94}.
Also, observations  showed that corona opacity is small \citep{Zdziarski00,Molina09}.
Therefore, the  projection effect should be rather weak in X-ray corona emission.
Light bending effect caused by the strong gravity around the SMBH could also produce anisotropy \citep{Fabian03, Chen13} in X-ray emission, however in the way contrary to our finding.

There are other factors that may control the relative strength of X-ray emission in AGNs, i.e., luminosity and Eddington ratio (normalized accretion rate), in the sense that AGNs with higher luminosity and/or Eddington ratio tend to have weaker X-ray emission relative to the accretion power \citep[e.g.][]{Fanali13}.
Our analyses are performed normalized to \Oline{IV} luminosity, thus have corrected the possible effect of luminosity.
Sy1s and Sy2s generally show consistent physical parameters, including black hole mass, luminosity and Eddington ratio \citep{Singh11}.
X-ray spectral slopes $\Gamma$ also measures the Eddington ratio \citep{Fanali13}.
Our spectral fittings yield consistent $\Gamma$ between Sy1s and Sy2s (Fig. \ref{res_Gamma}), indicating  that two subsamples have on average the same Eddington  ratio.
Therefore the observed relatively stronger intrinsic hard X-ray emission in Sy1s than in Sy2s are not caused by different Eddington ratios between the two subsamples.  

Under the scheme of AGN unification model, the difference we find between Sy1 and Compton-thin Sy2 indicates anisotropy of  the X-ray emission, which can be attributed to the beaming effect  of corona outflowing along the axis of the accretion system.

\begin{figure}
\centering
\plotone{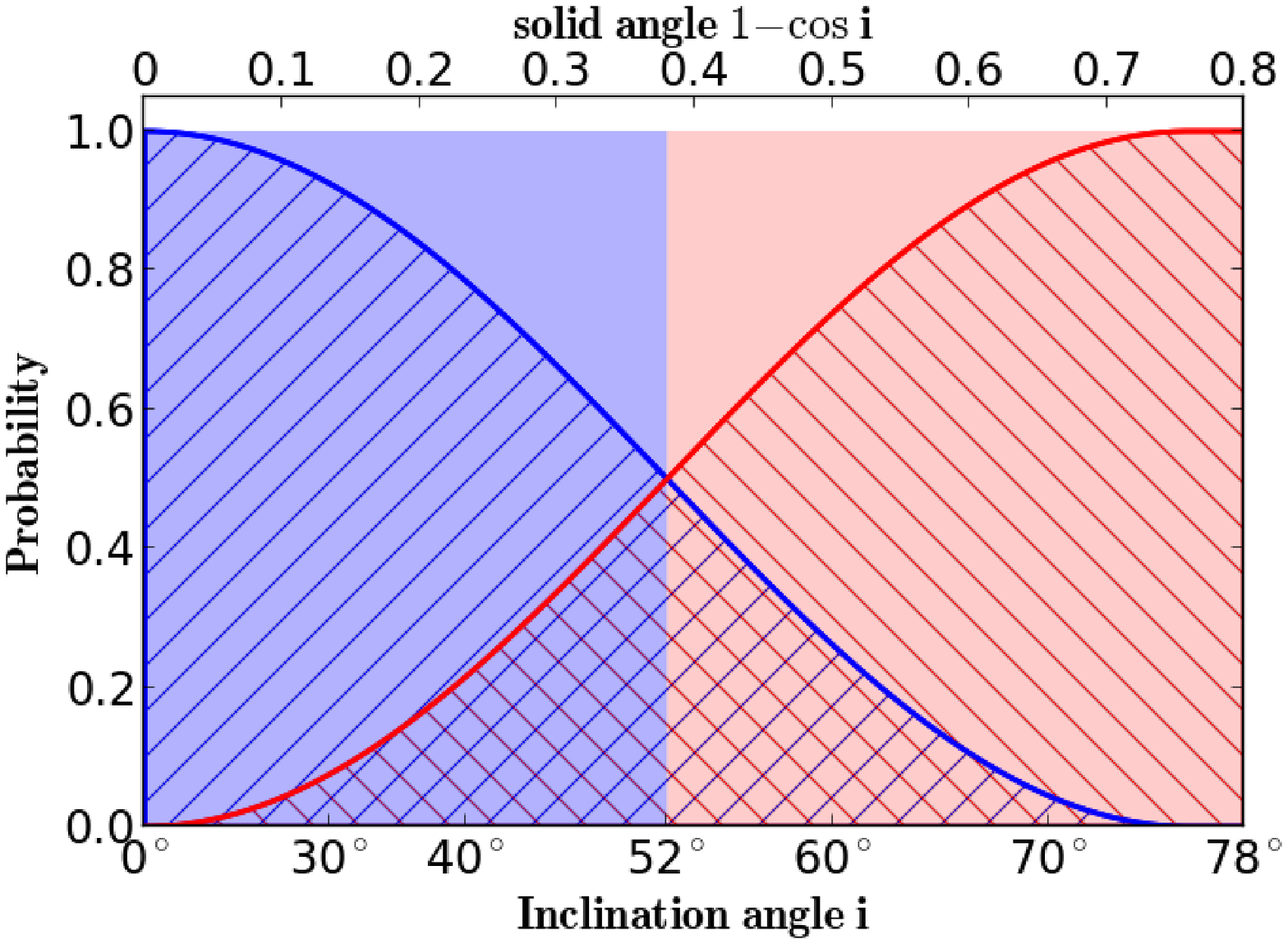}
\plotone{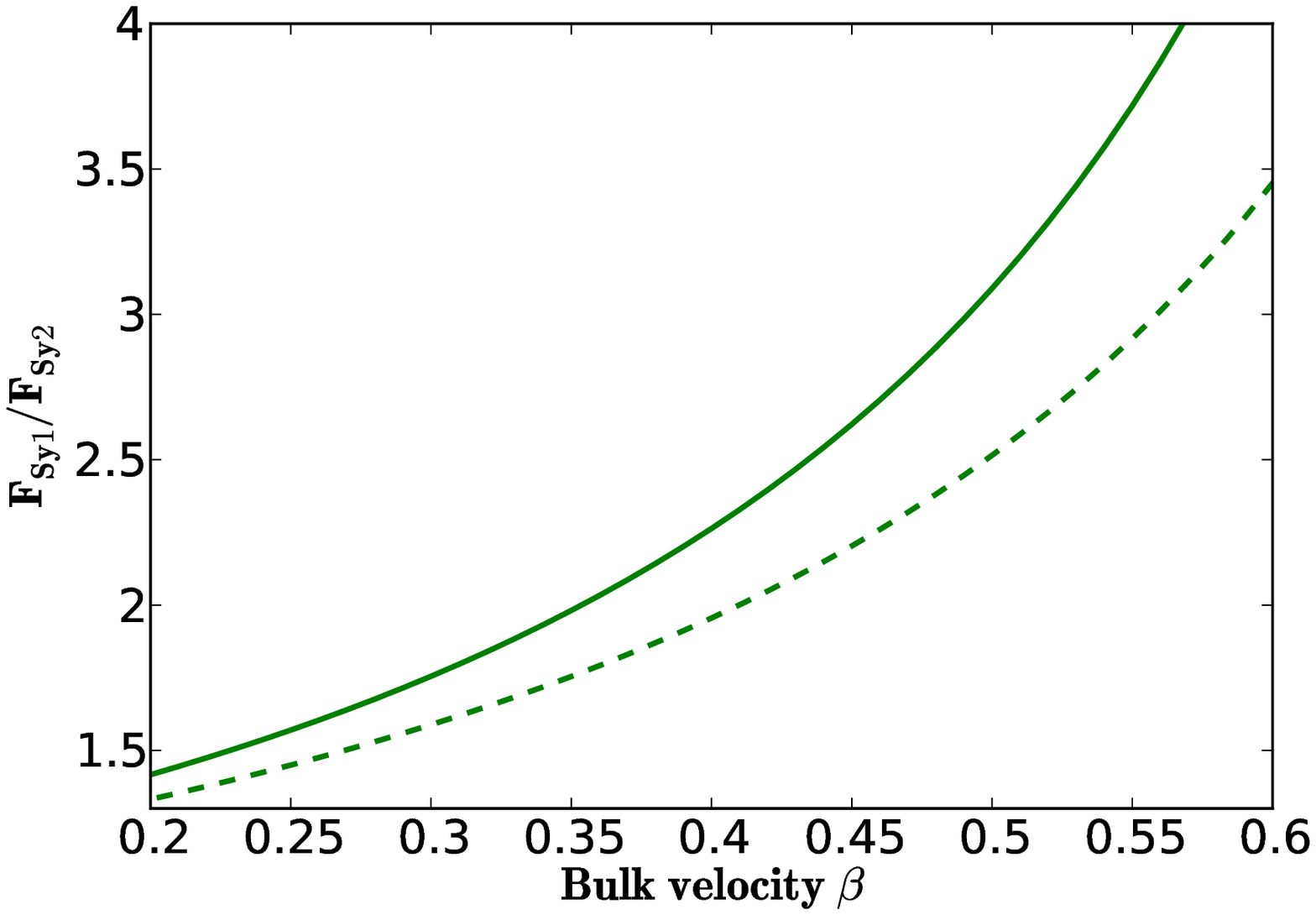}
\caption{In the upper panel, we plot the probability that a Seyfert galaxy could appear as Sy1 or Sy2 as a function of the inclination angle $i$ (or the solid angle 1 - cos $i$) in our toy models.
The filled blue and red regions mark the distinct inclination angle ranges of $0\arcdeg-52\arcdeg$ for Sy1s and $52\arcdeg-78\arcdeg$ for Compton-thin Sy2s.
With these assumptions, the required bulk outflowing velocity to explain the average difference in the intrinsic X-ray emission between Sy1s and Compton-thin Sy2s ($F_{Sy1}$/$F_{Sy2}$) is plotted with a solid line in the lower panel.
The hachured regions under the blue (Sy1) and red (Sy2) curves in the upper panel plot a comparison model. In such a toy model, the probability of a Seyfert galaxy being viewed as a Sy1 (blue curve) is a cosine function of the solid angle, which reduces from 1 at $0\arcdeg$ to 0.5 at $52\arcdeg$, in the sense that at the otherwise assumed transition angle of $i$ = $52\arcdeg$ between Sy1s and Sy2s, a Seyfert galaxy has equal possibility to appear as a Sy1 or a Sy2.
Under this situation, the required bulk outflowing velocity is higher, as shown by the dashed line in the lower panel.
\label{Sy1Sy2overlapping}}
\end{figure}

\begin{figure}
\centering
\plotone{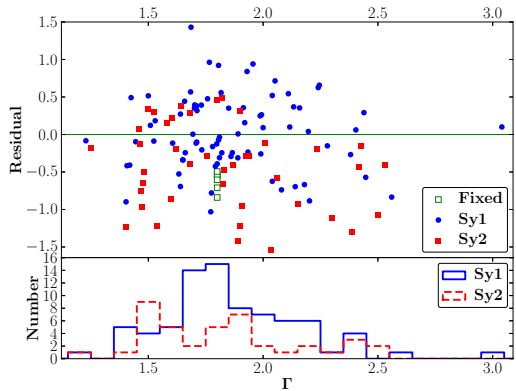}
\caption{The residuals in Fig. 1 as a function of X-ray spectral photon index $\Gamma$ (upper panel), which is a nice indicator of Eddington ratio in AGNs.
The trend of weaker X-ray emission in Sy2s is visible from low to high $\Gamma$ values.
For a couple of sources with too few X-ray counts, $\Gamma$ is fixed at 1.8 (green squares).
The median (mean) $\Gamma$ of Sy1s and Sy2s are 1.81 (1.86) and 1.82 (1.83) respectively.
K-S test also shows there is no statistical difference between  the two distributions of $\Gamma$ (lower panel).
\label{res_Gamma}}
\end{figure}

Below we calculate the bulk outflowing velocity of the corona required to produce the observed anisotropy.
Under the scheme of AGN unification model, we attribute the classification of Sy1, Compton-thin Sy2 and Compton-thick Sy2 to different ranges of inclination angles.
Based on  study of the $12\micron$ galaxy sample, an obscured fraction of AGN of 62\% and a Compton-thick fraction of AGN of $\sim$20\% were reported \citep{Brightman11b}.
According to these values, we define $0\arcdeg-52\arcdeg$ and $52\arcdeg-78\arcdeg$ as the ranges of inclination angles of Sy1s and Compton-thin Sy2s (see Fig. \ref{Sy1Sy2overlapping}).

Assuming an optical-thin corona with an outflowing velocity of $\beta = v/c$, the beaming factor is defined as the relative intensity observed from an inclination angle $i$:
\[b = (\gamma^{-1}(1-\beta\cos i)^{-1})^{3+\alpha}\]
where $\gamma=(1-\beta^2)^{-1/2}$, $\alpha$ is the energy index of the power-law.
Here a typical value of $\alpha=0.8$ is taken.
We calculate the ratio of the beaming factors averaged within the ranges of Sy1s and Compton-thin Sy2s at different bulk velocities, and we find that a flux decrement of $2.8^{+0.5}_{-0.4}$ times of Compton-thin Sy2s comparing  with Sy1s indicates an outflowing velocity of $0.47\pm0.05$ $c$ (see Fig. \ref{Sy1Sy2overlapping}).
This corresponds to  the escape speed at $\sim 7-12\ GM/c^2$, and  the electron thermal speed in plasma with temperature T $\sim 64$ keV (E = kT). 
The factor of $1.70_{-0.25}^{+0.30}$ (if simply comparing sources with $N_H$ $>$ 10$^{22}$ cm$^{-2}$  with those with smaller $N_H$, regardless of their optical identifications, see \S \ref{sampleincompletenessbias}) can be translated to an outflowing velocity of
$0.29_{-0.08}^{+0.06}$ $c$, and
the flux decrement of $2.2^{+0.9}_{-0.3}$ in 14 -- 195 keV emission (see Fig. \ref{14-195-OIV}) corresponds to an outflowing velocity of $0.38_{-0.06}^{+0.12}$ $c$. 

The difference in inclination angle between Sy1s and Sy2s is likely not as distinct as we have assumed, as suggested by the revised unification model \citep{Elitzur12}, the clumpy torus model \citep{Nenkova08}, the possibility that some Sy2s are due to large scale  obscurations in the host  galaxies \citep{Matt00,Malkan98,Rutkowski13}, and the misalignment between the axis of jet and the axis of obscuration in some AGNs \citep{Raban09}.
For example, the clumpy torus model could allow for Sy1s viewed edge-on and Sy2s viewed face-on, although with very low probabilities. 
These effects would reduce the difference in  inclination between two populations (Sy1s and Sy2s), and require even higher outflowing velocities.
To illustrate such effect, we adopt a simple toy model by assuming that the probability of a Seyfert galaxy appearing as Sy1 gradually reduces from face-on to edge-on (as a cosine function of the solid angle, see the upper panel of Fig. \ref{Sy1Sy2overlapping} and the caption for details), and compare the outflowing velocity required in this case with that in the above case (see the lower panel of Fig. \ref{Sy1Sy2overlapping}).
More accurate calculations rely on future exact knowledge on the inclinations of Sy1s and Compton-thin Sy2s.

In radio-loud AGNs which constitute only $\sim 10-20\%$ of the AGN population \citep{Kellermann89}, the radio emission originates from powerful relativistic jets up to kpc--Mpc scales, the launch of which is an outstanding puzzle in astrophysics \citep{Nemmen12}.
In radio-quiet ones, jet-like radio morphologies are also often detected \citep{Panessa13}, though much weaker and more compact.
Learning how weak and compact jets are launched can help to  understand the nature of more powerful ones.
Strong correlations between X-ray and radio emission were detected in radio-quiet AGNs \citep{Panessa07,Laor08,Wu13} and in X-ray binaries \citep{Merloni03}, suggesting a close corona-jet coupling.
How could this happen is still unclear.
Popular theoretical interpretations include: both X-ray and core radio emission originate from static (non-outflowing) coronal activity \citep{Laor08}; magnetically dominated corona produces jet/outflow that is responsible for radio emission \citep{Merloni02}; aborted jets \citep{Ghisellini04} produce both radio and X-ray emission; or the role of X-ray corona itself could be subsumed by the jet base \citep{Markoff05}.
Our finding remarkably matches the scenario that the role of corona could be subsumed by the base of  jet in AGNs \citep{Markoff05}, and could naturally explain the mysterious correlation between radio and X-ray emission in radio-quiet AGNs.
Two outstanding puzzles in astrophysics, the launches of coronae and relativistic jets, could therefore be directly bound, and even be merged.
The same picture could be applied to other black hole accretion systems, such as X-ray binaries.

\subsection{Final Remarks}
\label{finalremarks}
We have found that X-ray corona emission is intrinsically weaker in type 2 AGNs than in type 1 AGNs.
Studies taking hard X-ray emission as isotropic luminosity proxy to calibrate emission in other bands could therefore have been biased.
This also implies that hard X-ray surveys (even at $> 10$ keV) are biased against not only Compton-thick AGNs, but also Compton-thin type 2 sources, and the obscured fraction of AGNs yielded from hard X-ray surveys could have been underestimated. 
Interestingly, by taking the anisotropy of hard X-ray emission (although interpreted differently) into account, one can tackle the inconsistence in the observed obscured AGN fraction and its dependence to luminosity between X-ray samples and optical, radio and IR selected ones \citep{Zhang05,Mayo13}, providing an independent support to the anisotropy of corona X-ray emission.

Outflowing coronae may produce systematically higher cut-off energies in the hard X-ray spectra in type 1 AGNs than in type 2 sources due to Doppler shift.
Assuming the viewing angles to type 1 AGNs span from $0\arcdeg - 52\arcdeg$ and $52\arcdeg - 90\arcdeg$ to type 2 AGNs \citep{Brightman11b}, the expected difference in cut-off energy between two populations is by a factor of 1.4, adopting a bulk outflowing velocity of $0.47c$.
This difference is however too small to be detected with current observations considering the large uncertainties and huge intrinsic scatter in the measured cut-off energies \citep{Molina13}.
Nevertheless, it could be an efficient approach to measuring the corona outflow in the future.
\citet{Ricci11} compared the composite INTEGRAL  spectrum of Sy2s with  that of Sy1s.
Interestingly, from their Fig. 7  we can see a likely drop at above 80 keV in the ratio between the composite spectra of Sy2s and Sy1s, suggesting Sy2s have lower cut-off  energies.
Although this pattern is statistically insignificant yet due to the limited data quality, this points a direction to future observations. 

Another consequence of bipolar outflowing corona model is relatively weaker reflection component from the accretion disk beneath the corona comparing with the static corona model  \citep{Beloborodov99,Miniutti10}, because of the relativistic beaming effect. 
An outflowing velocity of $0.4c$ would reduce the relative strength of disk reflection by a factor of $\sim$ 2 -- 8 comparing with a static corona \citep[ viewing inclination dependent,][]{Beloborodov99}.
However, the detection of strong reflection component from the accretion disk in some AGNs \citep[e.g.][]{Tanaka95,Reynolds97,Nandra97}, such as the broad \FeKa{} line emission, does not necessarily argue against the model of outflowing corona.  In  addition to the inclination angle effect, it is possible that the corona outflowing velocity in AGNs is not uniform.
It could be very low in some AGNs, producing no beaming effect, thus could explain (likely together with the light bending effect) the detected strong broad \FeKa{} lines.
Higher outflowing velocity leading to stronger beaming effect, could also naturally explain the non-detection of broad \FeKa{} lines in many AGNs \citep[e.g.][]{Nandra07}.
Alternatively, there could be both an outflowing corona, and a non-outflowing (or even inflowing like magnetic loops) corona \citep{Wilkins12}.
In this scheme, the direct observed continuum is dominated by the emission from the outflowing corona, while the disk is mainly illuminated by the non-outflowing (or inflowing) corona.
As the variations of the outflowing corona and the non-outflowing (or inflowing) corona do not necessarily correlate, this diagram may naturally explain the puzzling constant disk reflection strength while the continuum emission is highly variable  in a few sources \citep[e.g.][]{Fabian02_MCG-6-30-15}, 
and the non-uniform variation pattern of the disk reflection relative to X-ray continuum in different sources \citep[e.g.][]{Iwasawa96,Wang99,Wang01,Shu10_NGC2992}.

Theoretical calculations \citep{Malzac01} show that outflowing coronae produce hard X-ray spectra  which are much flatter than observed values in AGNs ($\Gamma \sim 1.4$ versus $\sim1.9$) with a bulk-outflowing velocity of $\sim 0.4c$.
However, this relies on the assumption that the disk reprocessed radiation is the main cooler of the coronal plasma, while soft radiation produced viscously from the accretion disk and bremsstrahlung/cyclo-synchrotron emission in the corona, have been neglected \citep{Malzac01}.
Such assumption could only be valid if the corona is in form of strong concentrated flares and locally the reprocessed flux from the disk dominates the soft seed radiation \citep{Haardt94}, and may not hold in reality.

We note the large dispersions in the correlation between X-ray and \Oline{IV} luminosities.
The standard  deviations of the residuals (as defined in \S \ref{statistic}, see Fig. \ref{2-10-OIV}) are 0.5 and 0.7 for Sy1s and Sy2s respectively, both (slightly) larger than the average difference  between Sy1s and Sy2s (log 2.8 $\sim$ 0.45). 
The large scatter could be caused by many factors, including X-ray variation, the structure of NLR, the spectral energy distribution of AGNs, etc.
Interestingly, a dispersion in the outflowing velocity of the corona in AGNs could also produce  a large scatter in the correlation between X-ray and \Oline{IV} luminosities. 
Further investigations are required to interpret both the large scatter and the offset in the correlations between Sy1s and Sy2s simultaneously.

\section{SUMMARY}
\label{sec:summary}

To investigate whether the corona emission in radio-quiet AGNs is isotropic, we  compile a large sample of AGNs with both \Oline{IV} 25.89 $\micron$ emission line flux measurements and X-ray spectra.
Only secularly identified Compton-thin sources are included.
Known radio-loud sources are also excluded.
Taking \Oline{IV}  luminosity as a proxy of AGN intrinsic luminosity, we compare the absorption-corrected 2 -- 10 keV (also \Swift-BAT 14 -- 195 keV) luminosities of Sy1s and Compton-thin Sy2s.
Our results are summarized as follows:
\begin{description}
\item[1]
At given \Oline{IV} luminosity, the absorption-corrected 2 -- 10 keV X-ray emission is stronger in Sy1s than in Compton-thin Sy2s at a confidence level $>99.9\%$ by a factor of $2.8^{+0.5}_{-0.4}$.
Consistent pattern is seen in
\Swift-BAT 14 -- 195 keV emission.

\item[2] Based on careful analyses, we argue that the difference can not be attributed to sample selection,
insufficient absorption corrections, or anisotropy in \Oline{IV} line emission.
This is the first solid detection of moderate anisotropy in AGN's corona X-ray emission.
\item[3]
Under the scheme of AGN unification model,  this result suggests the coronae in AGNs are bipolar outflowing with a velocity of $\sim 0.3 - 0.5c$.
Such outflowing coronae could naturally be linked to the  bases of weak jets in such systems.

\item[4]
 We discuss the implications of outflowing coronae, including the relative strength of disk reflection component (i.e. the broad \FeKa{} line), the selection bias in hard X-ray surveys, the validity of using hard X-ray emission as AGN luminosity proxy,  the obscured fraction of AGNs in X-ray surveys, and a particular prediction that Sy2s should have smaller X-ray power-law cutoff energies comparing with Sy1s, which could be testified with future hard X-ray observations.
\end{description}

\acknowledgments

\textbf{This work is supported by Chinese NSF through grant 10825312 \& 11233002.
JXW acknowledges support from Chinese Top-notch Young Talents Program.
We thank Tinggui Wang, Feng Yuan, Andrzej Zdziarski and Fuguo Xie for discussions, and the anonymous referee for helpful comments.} 

\appendix
\section{Short Notes on Individual Sources}
\label{Apd:collection}
A short note is given to each source whose intrinsic 2 -- 10 keV  flux is taken from literature.
Generally their X-ray spectra are of high quality and have been nicely interpreted in previous studies.
For sources with multiple observations, we adopt their exposure-time weighted mean intrinsic luminosities in logarithm space.

\begin{description}
\item[NGC3516] 
\citet{Turner05_NGC3516} performed detailed spectral analysis on three \XMM{}/\Chandra{} observations of NGC 3516, revealed three distinct absorbing components. 

\item[2MASX J05580206-3820043 (H 0557-385)] 
\citet{Longinotti09_H0557-385} found with \XMM{} data that 2MASX J05580206-3820043 is absorbed by both ionized gas and partially covering neutral gas.
It showed extreme flux variation by a factor of 10, entirely due to intervening line-of-sight clouds. 

\item[ESO 434- G 040 (MCG-5-23-16)] 
\citet{Braito07_MCG-5-23-16} revealed with \Newton-RGS spectrum that the soft X-ray emission of ESO 434-G040 is likely dominated by several emission lines superimposed on an unabsorbed scattered power-law continuum. 
This Sy1.9 has a broad \FeKa{} line.

\item[UGC 03973 (Mrk79)] 
\citet{Gallo11_Mrk79} fitted the spectra of UGC 03973 (\XMM{}/\Suzaku{}) with a power-law with ionized reflection and absorption, plus a broad \FeKa{} line. 

\item[NGC 3227] 
\citet{Markowitz09_NGC3227} modeled the time-averaged spectrum of NGC 3227 (observed by \XMM{}) as a moderately flat power-law with two ionized absorbers. 

\item[NGC 3783] 
\citet{Krongold03_NGC3783} analyzed the \Chandra{} HETGS spectrum of NGC 3783.
Their model consists of a two-phase ionized absorber. 

\item[NGC 4388] 
\citet{Beckmann04_NGC4388} presented {\it INTEGRAL} and \XMM{} observations of NGC 4388, and found heavy absorption (2.7 $\times$ 10$^{23}$ $cm^{-2}$).

\item[NGC 4395] 
\citet{Nardini11_NGC4395} analyzed the \XMM{} and \Suzaku{} spectra of the dwarf Seyfert NGC 4395, attributing its spectral variation to varying partial covering absorber. 

\item[NGC 4507] 
\citet{Braito13_NGC4507} presented the \Suzaku{} observation of NGC 4507, found a variable absorber and a strong reflected component. 
 
\item[ESO 323- G 077] 
\citet{Jimenez08_ESO323-G077} analyzed the \XMM{} spectra of ESO 323-G077, found a neutral absorber (with N$_H$ = 5.82 $\times$ 10$^{22}$ $cm^{-2}$), and two ionized absorber plus a broad \FeKa{} line. 

\item[NGC 1365] 
\citet{Risaliti09_NGC1365} detected an absorption with $N_H \sim 3.5 \times 10^{23} cm^{-2}$ crossing the line of sight based on a $60$ ks \XMM{} observation. A broad \FeKa{} line is also detected. 
\item[FAIRALL 0049 (IRAS 18325-5926)] 
 \citet{Mocz11_IRAS18325} detected a blue-shifted ionized absorber in the \Chandra{} HETGS spectrum of IRAS 18325-5926. 

\item[NGC 5506] 
\citet{Guainazzi10_NGC5506_broad} found a broad \FeKa{} line in the X-ray obscured NLS1 NGC 5506 using \XMM{} data. 

\item[MRK 0273] 
The ULIRG Mrk 273 is merging with an unabsorbed Sy2 Mrk 273x, which lies $1.2$ arcmin away. \citet{Balestra05_Mrk273} used three collision ally ionized plasma components in addition to an absorbed power-law to fit the \XMM{} spectrum of Mrk 273.
The intrinsic flux is $22.1\times10^{-13}$.

\item[MCG -03-34-064] 
\citet{Miniutti07_IRAS13197} analyzed the \XMM{} spectrum of MCG-3-34-64 (IRAS 13197-1627), recovering Compton-thin absorption and a broad \FeKa{} emission line. 

\item[NGC 4151] 
\citet{Lubinski10_NGC4151} present a comprehensive spectral analysis of NGC 4151, using the {\it INTEGRAL}, {\it RXTE}, \XMM{}, {\it Swift} and {\it Suzaku}, considering ionized absorber, partial covering absorber, and reflection.  

\item[MRK 0463E] 
\citet{Bianchi08_Mrk463_biAGN} analyzed the \Chandra{}, \XMM{} and {\it HST} data of the double nucleus ULIRG Mrk 463, which is consist of Mrk 463E and Mrk 463W with a projected separation of 3.83\arcsec.
Their 2 -- 10 keV intrinsic luminosities are $1.5\times10^{-43}$ and $3.8\times10^{-42}\ erg\ s^{-1}$ respectively, as measured with \Chandra{} data.
The \Oline{IV} flux of Mrk 463E is measured with \Spitzer{} IRS LH module (slit width 11.1\arcsec), thus may be polluted by Mrk 463W.

\item[NGC 4253 (Mrk 766)] 
The NLS1 Mrk 766 is highly variable in the hard X-ray flux on time-scales as short as a few hundred seconds. \citet{Turner07_Mrk766} attributes the spectral variability of Mrk 766 to the variations in a complex and multi-layered absorber. 

\item[NGC 6860] 
\citet{Winter10_NGC6860} analyzed the \XMM{} and \Suzaku{} observations of NGC 6860,  and found a two-component warm ionized absorber and a broad \FeKa{} line. 

\item[NGC 0788, NGC 6300, NGC 7172, ESO 103- G 035] 
The 2 -- 10 keV intrinsic luminosities of these sources are taken from the broad-band X-ray spectral analysis by \citet{DeRosa12} using various available instruments including \XMM{}, \Chandra{}, \Suzaku{}, \INTEGRAL{} and \BeppoSAX{}.

\item[FAIRALL 0009] 
\citet{Lohfink12_Fariall9} fitted the multi-epoch spectra of FAIRALL 9 as observed by \XMM{} and \Suzaku{} with a model including a relativistically ionized reflection component.

\item[NGC 1052] 
\citet{Brenneman09_NGC1052} analyzed the 101 ks Suzaku spectrum of NGC 1052.
The 0.5--10 keV continuum is well modeled by a power-law continuum modified by Galactic and intrinsic absorption and a soft, thermal component below 1 keV.
Broad \FeKa{} line is also detected.

\item[MRK 335] 
MRK 335 was monitored by \XMM{} with 200 ks exposure.
Its spectra show warm absorber and blurred reflection \citep{Gallo13_Mrk335}.
\item[UGC 05025] 
\citet{Gallo05_Mrk705} fitted the \Chandra{} spectrum of UGC 05025 (MRK 705) with a primary power-law and an additional broken power-law representing the soft excess.

\item[MRK 1298] 
\citet{Giustini11_Mrk1298} found massive ionized absorbers along the line of sight of MRK 1298 (PG 1126-041).
\item[IC 5063] 
\citet{LaMassa11} fitted the \Chandra{} spectrum of IC 5063 with a double absorbed power-law.

\item[ESO 383-G035] 
\citet{Miller08_MCG-6-30-15} compiled all the available long-exposure, high-quality data for ESO 383-G035 (MCG-6-30-15): 522 ks of \Chandra{} HETGS, 282 ks of \XMM{} PN/RGS and 253 ks of \Suzaku{} XIS/PIN data.
A variable partial-covering zone plus absorbed low-ionization reflection, distant from the source, provides a complete description of the variable X-ray spectrum. 
\item[NGC 1097] 
With the high resolution observation of \Chandra{}, \citet{Nemmen06_NGC1097} eliminated the nearby ULX and surrounding star-forming ring of NGC 1097, and presented the spectrum of the nucleus region. 
\item[IC 4329A] 
\citet{Steenbrugge05_IC4329a} detected seven distinct absorbing systems in the high-resolution X-ray spectrum of IC 4329A taken with \XMM{}. 
\item[NGC 4051] 
\citet{Steenbrugge09_NGC4051} fitted the \Chandra{} LETGS high-resolution spectrum of NGC 4051 with multiple warm absorber models. 
\item[NGC 1566, NGC 4941] 
\citet{Taiki13_NGC1566_NGC4941} presented broad band (0.5--195 keV) X-ray spectra of NGC 1566 and NGC 4941 observed with both \Suzaku{} and \Swift{} BAT. 

\end{description}

\clearpage
\LongTables
\begin{deluxetable*}{lrrclclll}
\tabletypesize{\scriptsize}
\tablecaption{\label{tab:data}The sample
}
\centering
\tablewidth{0pt}
\tablehead{
\colhead{Name} &\colhead{Ra} &\colhead{Dec} &\colhead{z} &\colhead{Type} &\colhead{Origin} &\colhead{$\log L_{[OIV]}$} &\colhead{$\log L_{2-10}$} &\colhead{$\log L_{14-195}$}
}
\startdata
MRK 0335                 &1.581333 &20.202917 &0.025785	&$1$	&1,3,4,5 &$41.089_{-0.012}^{+0.012}$	&$42.750^{(1)}$	&$43.447_{-0.704}^{+0.346}$\\
NGC 0262                 &12.196417 &31.956972 &0.015034	&$2^{(2)}$	&1,3,5 &$40.957_{-0.008}^{+0.008}$	&$43.324_{-0.032}^{+0.033}$	&$43.899_{-1.573}^{+0.307}$\\
UGC 00545                      &13.395583 &12.693389 &0.058900	&$1$	&5 &$41.352$	&$43.645_{-0.003}^{+0.003}$	&$44.048u$\\
MCPS J00.898951-70.63450 &13.483958 &-70.634500 &0.068900	&$2^{(3)}$	&1,5 &$41.999_{-0.018}^{+0.019}$	&$43.663_{-0.026}^{+0.028}$	&$44.190u$\\
MRK 0352                 &14.972000 &31.826917 &0.014864	&$1$	&3 &$39.111_{-0.073}^{+0.062}$	&$42.758_{-0.007}^{+0.007}$	&$43.173_{-0.860}^{+0.331}$\\
NGC 0454                 &18.593875 &-55.398694 &0.012158	&$2^{(4)}$	&3 &$40.718_{-0.024}^{+0.023}$	&$42.308_{-0.072}^{+0.070}$	&$42.761_{-0.655}^{+0.350}$\\
FAIRALL 0009             &20.940750 &-58.805778 &0.047016	&$1$	&3,4 &$41.511_{-0.023}^{+0.022}$	&$43.782^{(5)}$	&$44.412_{-1.153}^{+0.316}$\\
NGC 0526A                &20.976625 &-35.065528 &0.019097	&$1$	&1,3,5 &$41.184_{-0.017}^{+0.016}$	&$43.266_{-0.003}^{+0.003}$	&$43.717_{-1.177}^{+0.316}$\\
NGC 0513                 &21.111875 &33.799444 &0.019544	&$2^{(6)}$	&1,3,5 &$40.755_{-0.029}^{+0.030}$	&$42.639_{-0.025}^{+0.023}$	&$43.245_{-0.695}^{+0.346}$\\
2MASX J01500266-0725482  &27.511250 &-7.430139 &0.017666	&$2^{(7)}$	&1,5 &$40.660_{-0.020}^{+0.021}$	&$41.757_{-0.022}^{+0.023}$	&$42.975u$\\
MRK 1014                 &29.959208 &0.394611 &0.163110	&$1$	&4,5 &$42.964_{-0.022}^{+0.021}$	&$43.818_{-0.032}^{+0.032}$	&$44.994u$\\
NGC 0788                 &30.276875 &-6.815528 &0.013603	&$2^{(8)}$	&2,3,5 &$40.990_{-0.007}^{+0.006}$	&$42.800^{(8)}$	&$43.522_{-1.253}^{+0.313}$\\
NGC 0863                 &33.639833 &-0.766694 &0.026385	&$1.2$	&3,4 &$40.574_{-0.058}^{+0.051}$	&$43.004_{-0.004}^{+0.004}$	&$43.424_{-0.575}^{+0.361}$\\
ESO 545- G 013$^{\dagger}$  &36.169250 &-19.142028 &0.033730	&$1$	&1,5 &$41.491_{-0.006}^{+0.006}$	&$42.515_{-0.067}^{+0.065}$	&$43.547u$\\
NGC 0931                 &37.060333 &31.311667 &0.016652	&$1$	&1,3,5 &$41.427_{-0.005}^{+0.005}$	&$43.285_{-0.003}^{+0.003}$	&$43.578_{-1.170}^{+0.316}$\\
NGC 973$^{\ddagger}$  &38.583792 &32.505611 &0.016195	&$2^{(9)}$	&3 &$40.921_{-0.090}^{+0.074}$	&$42.624_{-0.035}^{+0.033}$	&$43.266_{-0.813}^{+0.335}$\\
NGC 1052$^{\dagger}$  &40.270000 &-8.255778 &0.005037	&$2^{(10)}$	&3 &$39.124_{-0.019}^{+0.018}$	&$41.685^{(10)}$	&$42.218_{-0.815}^{+0.335}$\\
NGC 1097$^{\dagger}$  &41.579375 &-30.274889 &0.004240	&$1$	&1,2,5 &$39.345_{-0.050}^{+0.056}$	&$40.839^{(11)}$	&$41.726u$\\
NGC 1144                 &43.800833 &-0.183556 &0.028847	&$2^{(2)}$	&1,5 &$41.006_{-0.014}^{+0.015}$	&$43.448_{-0.037}^{+0.045}$	&$44.228_{-1.257}^{+0.313}$\\
ESO 417- G 006           &44.089750 &-32.185583 &0.016291	&$2^{(12)}$	&3 &$40.383_{-0.020}^{+0.019}$	&$42.518_{-0.022}^{+0.021}$	&$43.263_{-0.906}^{+0.328}$\\
MCG -02-08-039           &45.127458 &-11.415722 &0.029894	&$2^{(6)}$	&1,5 &$41.465_{-0.007}^{+0.007}$	&$42.680_{-0.091}^{+0.090}$	&$43.440u$\\
MRK 0609                 &51.355667 &-6.143861 &0.034488	&$1.8^{(13)}$	&4 &$41.356_{-0.162}^{+0.118}$	&$42.536_{-0.035}^{+0.027}$	&$43.567u$\\
NGC 1365                 &53.401542 &-36.140389 &0.005457	&$1$	&1,2,3,5 &$41.120_{-0.029}^{+0.030}$	&$42.253^{(14)}$	&$42.626_{-1.259}^{+0.313}$\\
NGC 1566$^{\ast}$  &65.001750 &-54.937806 &0.005017	&$1.5$	&5 &$39.670$	&$41.193^{(15)}$	&$42.076_{-0.767}^{+0.339}$\\
MRK 0618                 &69.092667 &-10.376056 &0.035550	&$1$	&1,5 &$41.471_{-0.024}^{+0.025}$	&$43.435_{-0.014}^{+0.013}$	&$43.716_{-0.654}^{+0.351}$\\
MCG -01-13-025           &72.922833 &-3.809250 &0.015894	&$1.2$	&3 &$39.636_{-0.117}^{+0.092}$	&$42.709_{-0.013}^{+0.012}$	&$43.248_{-0.789}^{+0.337}$\\
ARK 120                  &79.047583 &-0.149833 &0.032713	&$1$	&3,4 &$40.979_{-0.045}^{+0.040}$	&$43.927_{-0.024}^{+0.020}$	&$44.238_{-1.170}^{+0.316}$\\
ESO 362- G 018           &79.899167 &-32.657583 &0.012445	&$1$	&1,3,5 &$40.508_{-0.014}^{+0.014}$	&$42.468_{-0.013}^{+0.014}$	&$43.233_{-1.103}^{+0.318}$\\
2MASX J05210136-2521450  &80.255792 &-25.362611 &0.042563	&$2^{(16)}$	&4,5 &$42.027_{-0.067}^{+0.058}$	&$43.301_{-0.028}^{+0.028}$	&$43.725_{-0.523}^{+0.369}$\\
2MASX J05580206-3820043  &89.508333 &-38.334639 &0.033870	&$1$	&1,3,5 &$41.096_{-0.046}^{+0.043}$	&$43.907^{(17)}$	&$43.890_{-0.981}^{+0.324}$\\
IC 0450                  &103.051042 &74.427083 &0.018813	&$1$	&1,3,5 &$41.586_{-0.002}^{+0.002}$	&$43.080_{-0.003}^{+0.003}$	&$43.686_{-1.194}^{+0.315}$\\
UGC 03973                &115.636667 &49.809639 &0.022189	&$1$	&1,3,5 &$41.665_{-0.022}^{+0.021}$	&$43.255^{(18)}$	&$43.717_{-1.069}^{+0.320}$\\
IC 0486                  &120.087417 &26.613528 &0.026875	&$1$	&3 &$41.267_{-0.011}^{+0.011}$	&$42.808_{-0.009}^{+0.009}$	&$43.726_{-0.815}^{+0.335}$\\
PG 0804+761              &122.744167 &76.045139 &0.100000	&$1$	&4 &$41.723_{-0.065}^{+0.056}$	&$44.273_{-0.005}^{+0.005}$	&$44.574_{-0.694}^{+0.347}$\\
2MASX J08244333+2959238  &126.180375 &29.989889 &0.025325	&$2^{(19)}$	&4 &$41.088_{-0.011}^{+0.011}$	&$42.655_{-0.077}^{+0.082}$	&$43.293u$\\
NGC 2622                 &129.545583 &24.895278 &0.028624	&$1.9_{(13)}^{(13)}$	&4 &$41.225_{-0.018}^{+0.017}$	&$42.322_{-0.480}^{+0.088}$	&$43.401u$\\
VII Zw 244                     &131.188792 &76.886028 &0.131000	&$1$	&5 &$41.773$	&$43.455_{-0.055}^{+0.054}$	&$44.786u$\\
PG 0844+349                    &131.926958 &34.751222 &0.064000	&$1$	&5 &$41.172$	&$43.653_{-0.005}^{+0.005}$	&$44.123u$\\
NGC 2685                 &133.894625 &58.734389 &0.002945	&$2^{(20)}$	&5 &$37.915$	&$39.747_{-0.168}^{+0.164}$	&$41.409u$\\
NGC 2655                 &133.907208 &78.223083 &0.004670	&$2^{(21)}$	&2,5 &$39.479_{-0.111}^{+0.088}$	&$41.156_{-0.092}^{+0.094}$	&$41.812_{-0.558}^{+0.363}$\\
MRK 0704                 &139.608375 &16.305333 &0.029234	&$1$	&1,4,5 &$41.368_{-0.015}^{+0.016}$	&$43.309_{-0.004}^{+0.004}$	&$43.813_{-0.906}^{+0.328}$\\
MCG -01-24-012           &140.192708 &-8.056139 &0.019644	&$2^{(22)}$	&3 &$40.944_{-0.035}^{+0.032}$	&$43.160_{-0.024}^{+0.025}$	&$43.552_{-0.996}^{+0.323}$\\
CGCG 121-075             &140.929167 &22.909056 &0.032349	&$1$	&4 &$41.314_{-0.010}^{+0.009}$	&$43.614_{-0.011}^{+0.011}$	&$43.984_{-0.961}^{+0.325}$\\
UGC 05025$^{\dagger}$  &141.513708 &12.734333 &0.029150	&$1$	&4 &$41.052_{-0.017}^{+0.016}$	&$43.654^{(23)}$	&$43.491_{-0.602}^{+0.357}$\\
NGC 2992                 &146.425208 &-14.326389 &0.007710	&$1.9^{(24)}$	&2,3,5 &$41.208_{-0.025}^{+0.024}$	&$41.938_{-0.021}^{+0.004}$	&$42.550_{-0.752}^{+0.340}$\\
ESO 434- G 040           &146.917333 &-30.948722 &0.008486	&$1.9_{(25)}^{(25)}$	&3 &$40.651_{-0.138}^{+0.105}$	&$43.170^{(25)}$	&$43.508_{-1.680}^{+0.306}$\\
MRK 1239                 &148.079583 &-1.612083 &0.019927	&$1$	&1,5 &$41.152_{-0.009}^{+0.010}$	&$42.276_{-0.132}^{+0.160}$	&$43.081u$\\
MESSIER 081              &148.888208 &69.065306 &-0.000113	&$1.5$	&5 &$37.844$	&$40.225_{-0.003}^{+0.003}$	&$40.379_{-0.676}^{+0.349}$\\
PG 1001+054                    &151.083917 &5.216806 &0.161076	&$1$	&5 &$41.571$	&$42.891_{-0.093}^{+0.088}$	&$44.982u$\\
NGC 3227                 &155.877417 &19.865056 &0.003859	&$1.5$	&2,3,4,5 &$40.344_{-0.014}^{+0.013}$	&$42.070^{(26)}$	&$42.558_{-1.448}^{+0.309}$\\
NGC 3516                 &166.697875 &72.568583 &0.008836	&$1$	&1,2,3,4,5 &$40.902_{-0.004}^{+0.004}$	&$42.505^{(27)}$	&$43.313_{-1.517}^{+0.308}$\\
MRK 1298                       &172.319417 &-4.402111 &0.061960	&$1$	&5 &$42.171$	&$43.312^{(28)}$	&$44.094u$\\
NGC 3783                 &174.757333 &-37.738667 &0.009730	&$1.2$	&2,3,4,5 &$40.916_{-0.007}^{+0.007}$	&$43.092^{(29)}$	&$43.582_{-1.575}^{+0.307}$\\
UGC 06728                &176.316750 &79.681500 &0.006518	&$1.2$	&3 &$39.638_{-0.084}^{+0.070}$	&$41.753_{-0.016}^{+0.017}$	&$42.409_{-0.935}^{+0.327}$\\
MRK 1457                 &176.840042 &52.449583 &0.048572	&$2^{(19)}$	&4 &$41.195_{-0.041}^{+0.038}$	&$42.492_{-0.248}^{+0.244}$	&$43.874u$\\
NGC 4051                 &180.790042 &44.531333 &0.002336	&$1.5_{(21)}$	&1,2,3,4,5 &$39.727_{-0.018}^{+0.017}$	&$41.129^{(30)}$	&$41.677_{-1.141}^{+0.317}$\\
NGC 4138                 &182.374083 &43.685306 &0.002962	&$1.9^{(21)}$	&2,3 &$38.594_{-0.040}^{+0.037}$	&$41.180_{-0.021}^{+0.020}$	&$41.764_{-0.916}^{+0.328}$\\
NGC 4151                 &182.635750 &39.405722 &0.003319	&$1.5$	&2,3,4,5 &$40.742_{-0.007}^{+0.007}$	&$42.595^{(31)}$	&$43.117_{-2.179}^{+0.302}$\\
NGC 4168                 &183.071958 &13.205194 &0.007388	&$1.9^{(21)}$	&2 &$39.227_{-0.165}^{+0.119}$	&$39.886_{-0.061}^{+0.058}$	&$42.211u$\\
PG 1211+143                    &183.573583 &14.053694 &0.080900	&$1$	&5 &$41.590$	&$43.763_{-0.003}^{+0.003}$	&$44.337u$\\
NGC 4235                 &184.291167 &7.191583 &0.008039	&$1.2$	&2,3,4,5 &$39.698_{-0.076}^{+0.063}$	&$41.598_{-0.016}^{+0.013}$	&$42.669_{-0.872}^{+0.331}$\\
NGC 4253                 &184.610458 &29.812861 &0.012929	&$1$	&1,3,5 &$41.238_{-0.006}^{+0.006}$	&$42.523^{(32)}$	&$42.909_{-0.903}^{+0.329}$\\
MESSIER 106              &184.739583 &47.303972 &0.001494	&$1.9^{(33)}$	&2,4,5 &$38.579_{-0.046}^{+0.042}$	&$40.569_{-0.009}^{+0.010}$	&$41.034_{-0.807}^{+0.336}$\\
NGC 4388                 &186.444792 &12.662083 &0.008419	&$2^{(34)}$	&1,2,3,5 &$41.702_{-0.027}^{+0.025}$	&$43.202^{(34)}$	&$43.641_{-1.792}^{+0.305}$\\
NGC 4395                 &186.453583 &33.546917 &0.001064	&$1.5_{(35)}$	&2,3,5 &$38.271_{-0.014}^{+0.014}$	&$40.300^{(36)}$	&$40.797_{-0.871}^{+0.331}$\\
MESSIER 049              &187.444833 &8.000472 &0.003326	&$2^{(37)}$	&2 &$39.210u$	&$39.791_{-0.022}^{+0.024}$	&$41.515u$\\
MRK 0771                 &188.015000 &20.158111 &0.063010	&$1$	&4,5 &$41.419_{-0.062}^{+0.054}$	&$43.450_{-0.009}^{+0.010}$	&$44.111_{-0.541}^{+0.365}$\\
NGC 4507                 &188.902625 &-39.909250 &0.011801	&$2^{(38)}$	&2,3,4,5 &$41.040_{-0.039}^{+0.035}$	&$43.310^{(38)}$	&$43.768_{-1.561}^{+0.307}$\\
NGC 4565                 &189.086583 &25.987667 &0.004103	&$1.9^{(21)}$	&5 &$39.183$	&$39.975_{-0.046}^{+0.043}$	&$41.697u$\\
MESSIER 058              &189.431333 &11.818194 &0.005060	&$1.5_{(20)}$	&5 &$39.258$	&$41.331_{-0.009}^{+0.009}$	&$41.880u$\\
2MASX J12384342+0927362  &189.681000 &9.460167 &0.082902	&$2^{(19)}$	&4 &$42.002_{-0.018}^{+0.017}$	&$43.665_{-0.103}^{+0.102}$	&$44.295_{-0.446}^{+0.382}$\\
NGC 4593                 &189.914292 &-5.344250 &0.009000	&$1$	&1,2,3,5 &$40.510_{-0.050}^{+0.043}$	&$42.802_{-0.007}^{+0.007}$	&$43.201_{-1.254}^{+0.313}$\\
MESSIER 104              &189.997625 &-11.623056 &0.003416	&$1.9^{(39)}$	&5 &$38.791$	&$40.578_{-0.012}^{+0.012}$	&$41.538u$\\
NGC 4639                 &190.718292 &13.257389 &0.003395	&$1.5_{(21)}$	&2,5 &$38.593_{-0.142}^{+0.107}$	&$40.094_{-0.031}^{+0.031}$	&$41.532u$\\
PG 1244+026                    &191.646875 &2.369111 &0.048178	&$1$	&5 &$40.915$	&$43.200_{-0.003}^{+0.002}$	&$43.866u$\\
NGC 4748$^{\dagger}$  &193.051917 &-13.414722 &0.014630	&$1$	&1,5 &$41.596_{-0.001}^{+0.001}$	&$42.473_{-0.030}^{+0.029}$	&$42.821_{-0.484}^{+0.376}$\\
MRK 0231                       &194.059292 &56.873667 &0.042170	&$1$	&5 &$41.597u$	&$42.472_{-0.029}^{+0.027}$	&$43.747u$\\
NGC 4941$^{\ast}$  &196.054750 &-5.551611 &0.003696	&$2^{(15)}$	&1,2,5 &$39.945_{-0.005}^{+0.005}$	&$41.109^{(15)}$	&$41.790_{-0.646}^{+0.351}$\\
PG 1302-102                    &196.387542 &-10.555389 &0.278400	&$1.2$	&5 &$42.806$	&$44.883_{-0.019}^{+0.020}$	&$45.512u$\\
ESO 323- G 077           &196.608875 &-40.414667 &0.015014	&$1.2$	&3 &$41.104_{-0.011}^{+0.011}$	&$42.815^{(40)}$	&$43.224u$\\
PG 1307+085                    &197.445833 &8.330056 &0.155000	&$1.2$	&5 &$41.688$	&$44.101_{-0.015}^{+0.015}$	&$44.946u$\\
NGC 5005                 &197.734292 &37.059194 &0.003156	&$2^{(41)}$	&1,2,5 &$39.098_{-0.033}^{+0.036}$	&$40.022_{-0.042}^{+0.043}$	&$41.469u$\\
FBQS J131217.7+351521          &198.074042 &35.255889 &0.182923	&$1.2$	&5 &$41.668u$	&$43.803_{-0.017}^{+0.017}$	&$45.105u$\\
NGC 5033                 &198.364458 &36.593944 &0.002919	&$1.5$	&5 &$39.238$	&$40.912_{-0.011}^{+0.011}$	&$41.401u$\\
MCG -03-34-064           &200.601917 &-16.728472 &0.016541	&$1.8_{(42)}^{(42)}$	&1,3,4,5 &$41.837_{-0.023}^{+0.021}$	&$42.643^{(42)}$	&$43.275_{-0.844}^{+0.333}$\\
CGCG 218-007             &200.951875 &43.301194 &0.027276	&$2^{(19)}$	&4 &$41.453_{-0.006}^{+0.006}$	&$42.731_{-0.113}^{+0.106}$	&$43.359u$\\
ESO 383- G 035           &203.973792 &-34.295528 &0.007749	&$1$	&1,3,5 &$40.525_{-0.014}^{+0.014}$	&$42.415^{(43)}$	&$42.927_{-1.166}^{+0.316}$\\
NGC 5273                 &205.534750 &35.654222 &0.003549	&$1.5$	&5 &$39.134$	&$41.251_{-0.008}^{+0.008}$	&$41.587_{-0.543}^{+0.364}$\\
MRK 0273                 &206.175458 &55.886861 &0.037780	&$2^{(44)}$	&4,5 &$42.266_{-0.014}^{+0.014}$	&$42.866^{(44)}$	&$43.648u$\\
IC 4329A                 &207.330292 &-30.309444 &0.016054	&$1$	&1,3,4,5 &$41.802_{-0.009}^{+0.009}$	&$44.066^{(45)}$	&$44.227_{-1.758}^{+0.305}$\\
UGC 08823                &208.264375 &69.308222 &0.030451	&$1$	&4 &$41.341_{-0.015}^{+0.015}$	&$43.634_{-0.030}^{+0.022}$	&$43.922_{-1.053}^{+0.320}$\\
PG 1351+640              &208.315958 &63.762694 &0.088200	&$1$	&5 &$41.382u$	&$43.053_{-0.012}^{+0.012}$	&$44.417u$\\
MRK 0463E                &209.012042 &18.371972 &0.050000	&$2^{(46)}$	&4,5 &$42.599_{-0.008}^{+0.008}$	&$43.274^{(46)}$	&$43.900u$\\
NGC 5506                 &213.312042 &-3.207583 &0.006181	&$1$	&1,2,3,4,5 &$41.306_{-0.005}^{+0.005}$	&$43.007^{(47)}$	&$43.310_{-1.721}^{+0.305}$\\
PG 1411+442              &213.451375 &44.003889 &0.089600	&$1$	&4,5 &$41.465_{-0.078}^{+0.066}$	&$43.491_{-0.268}^{+0.239}$	&$44.431u$\\
NGC 5548                 &214.498042 &25.136778 &0.017175	&$1$	&1,3,4,5 &$40.992_{-0.026}^{+0.024}$	&$43.418_{-0.002}^{+0.002}$	&$43.719_{-1.309}^{+0.312}$\\
MRK 1383                 &217.277458 &1.285139 &0.086570	&$1$	&4,5 &$41.773_{-0.070}^{+0.060}$	&$44.204_{-0.016}^{+0.018}$	&$44.522_{-0.608}^{+0.357}$\\
UGC 09412                &219.091958 &58.794278 &0.031455	&$1$	&1,3,5 &$41.161_{-0.018}^{+0.018}$	&$43.492_{-0.011}^{+0.011}$	&$43.772_{-0.918}^{+0.328}$\\
MRK 0478                 &220.531083 &35.439694 &0.079055	&$1$	&4,5 &$41.939_{-0.021}^{+0.020}$	&$43.537_{-0.006}^{+0.006}$	&$44.316u$\\
PG 1448+273                    &222.786500 &27.157472 &0.065000	&$1$	&5 &$42.010$	&$43.296_{-0.010}^{+0.010}$	&$44.137u$\\
IC 4518A                       &224.421583 &-43.132111 &0.016261	&$2^{(8)}$	&5 &$41.704$	&$42.373_{-0.054}^{+0.063}$	&$43.218_{-0.814}^{+0.335}$\\
MRK 0841                 &226.005000 &10.437833 &0.036422	&$1.5$	&3,5 &$41.859_{-0.021}^{+0.020}$	&$43.492_{-0.003}^{+0.003}$	&$44.040_{-0.908}^{+0.329}$\\
2MASX J15115979-2119015  &227.999167 &-21.317139 &0.044607	&$1$	&1,5 &$42.162_{-0.004}^{+0.004}$	&$43.576_{-0.022}^{+0.015}$	&$44.166_{-0.737}^{+0.342}$\\
NGC 5899                 &228.763417 &42.049833 &0.008546	&$2^{(48)}$	&5 &$40.553$	&$42.187_{-0.022}^{+0.020}$	&$42.514_{-0.754}^{+0.340}$\\
NGC 5995$^{\ddagger}$  &237.103958 &-13.757778 &0.025194	&$2^{(49)}$	&1,3,5 &$41.265_{-0.023}^{+0.022}$	&$43.066_{-0.094}^{+0.129}$	&$43.801_{-0.893}^{+0.329}$\\
UGC 10120                &239.790125 &35.029861 &0.031328	&$1$	&4 &$40.808_{-0.080}^{+0.067}$	&$42.875_{-0.013}^{+0.013}$	&$43.482u$\\
MRK 0876                 &243.488250 &65.719333 &0.129000	&$1$	&4,5 &$42.401_{-0.021}^{+0.020}$	&$44.301_{-0.021}^{+0.020}$	&$44.737_{-0.575}^{+0.362}$\\
MRK 0877$^{\dagger}$  &245.047042 &17.407694 &0.112438	&$1$	&5 &$41.106$	&$44.016_{-0.023}^{+0.020}$	&$44.642u$\\
SBS 1626+554                   &246.983833 &55.375417 &0.133000	&$1$	&5 &$40.974u$	&$44.135_{-0.021}^{+0.020}$	&$44.800u$\\
NGC 6221                 &253.192000 &-59.218611 &0.004999	&$2^{(50)}$	&5 &$40.044$	&$41.173_{-0.022}^{+0.021}$	&$42.033u$\\
NGC 6300                 &259.247792 &-62.820556 &0.003699	&$2^{(8)}$	&5 &$39.971$	&$41.800^{(8)}$	&$42.478_{-1.348}^{+0.311}$\\
{[}HB89] 1821+643          &275.488792 &64.343444 &0.297000	&$1$	&4 &$43.836_{-0.023}^{+0.022}$	&$45.625_{-0.007}^{+0.006}$	&$45.731_{-0.795}^{+0.337}$\\
FAIRALL 0049             &279.242875 &-59.402389 &0.020021	&$2^{(51)}$	&3,4 &$41.551_{-0.021}^{+0.020}$	&$43.347^{(51)}$	&$43.109_{-0.603}^{+0.358}$\\
ESO 103- G 035           &279.584750 &-65.427556 &0.013286	&$2^{(8)}$	&3,4 &$41.112_{-0.016}^{+0.015}$	&$43.400^{(8)}$	&$43.645_{-1.427}^{+0.309}$\\
ESO 140- G 043           &281.224917 &-62.364833 &0.014178	&$1$	&3,4 &$41.071_{-0.007}^{+0.007}$	&$42.929_{-0.010}^{+0.009}$	&$43.260_{-0.983}^{+0.324}$\\
ESO 141- G 055           &290.308917 &-58.670306 &0.037109	&$1$	&1,5 &$41.367_{-0.007}^{+0.007}$	&$43.862_{-0.029}^{+0.024}$	&$44.241_{-1.094}^{+0.319}$\\
NGC 6814                 &295.669333 &-10.323500 &0.005214	&$1.5$	&5 &$40.211$	&$42.176_{-0.006}^{+0.006}$	&$42.667_{-1.160}^{+0.316}$\\
NGC 6860                 &302.195375 &-61.100194 &0.014884	&$1.5_{(52)}$	&1,3,5 &$40.774_{-0.011}^{+0.011}$	&$43.056^{(52)}$	&$43.423_{-1.073}^{+0.320}$\\
MRK 0509                 &311.040583 &-10.723472 &0.034397	&$1$	&1,2,3,4,5 &$41.833_{-0.010}^{+0.010}$	&$44.104_{-0.002}^{+0.002}$	&$44.421_{-1.266}^{+0.313}$\\
IC 5063$^{\dagger}$  &313.009750 &-57.068778 &0.011348	&$2^{(16)}$	&1,3,5 &$41.520_{-0.031}^{+0.028}$	&$42.941^{(16)}$	&$43.320_{-1.182}^{+0.315}$\\
UGC 11763                &323.115875 &10.138750 &0.062977	&$1$	&4,5 &$42.004_{-0.016}^{+0.016}$	&$43.521_{-0.008}^{+0.008}$	&$44.109u$\\
NGC 7172                 &330.507875 &-31.869667 &0.008683	&$2^{(8)}$	&1,2,3,4,5 &$40.843_{-0.022}^{+0.020}$	&$43.000^{(8)}$	&$43.457_{-1.505}^{+0.308}$\\
NGC 7213                 &332.317958 &-47.166611 &0.005839	&$1$	&1,2,3,5 &$39.288_{-0.105}^{+0.084}$	&$42.050_{-0.002}^{+0.002}$	&$42.502_{-0.965}^{+0.325}$\\
MRK 0304                       &334.301083 &14.239139 &0.065762	&$1$	&5 &$41.135$	&$43.592_{-0.017}^{+0.017}$	&$44.006_{-0.447}^{+0.384}$\\
NGC 7314                 &338.942458 &-26.050472 &0.004763	&$1$	&1,2,3,5 &$40.532_{-0.037}^{+0.033}$	&$42.339_{-0.004}^{+0.003}$	&$42.420_{-1.044}^{+0.321}$\\
MCG -03-58-007           &342.404792 &-19.274000 &0.031462	&$2^{(53)}$	&1,5 &$41.303_{-0.016}^{+0.017}$	&$42.644_{-0.104}^{+0.104}$	&$43.485u$\\
{[}HB89] 2251+113                &343.543333 &11.610639 &0.325500	&$1.2$	&5 &$43.040$	&$44.681_{-0.013}^{+0.012}$	&$45.675u$\\
NGC 7469                 &345.815083 &8.874000 &0.016317	&$1.2$	&2,3,4,5 &$41.330_{-0.051}^{+0.046}$	&$43.170_{-0.009}^{+0.009}$	&$43.602_{-1.184}^{+0.315}$\\
MRK 0926                 &346.181167 &-8.685722 &0.046860	&$1.5$	&3,4 &$41.834_{-0.043}^{+0.038}$	&$44.190_{-0.005}^{+0.005}$	&$44.771_{-1.376}^{+0.310}$\\
NGC 7603                 &349.735917 &0.243944 &0.029524	&$1.5$	&3 &$40.825_{-0.024}^{+0.023}$	&$43.544_{-0.007}^{+0.006}$	&$44.003_{-1.046}^{+0.321}$\\
\enddata
\tablecomments{ 
(1): name of object.
We mark the sources with neither obscuration-corrected fluxes found in literature nor usable \XMM{} spectra with flags representing the data origins: $^\dagger$ for \Chandra{}, $^\ddagger$ for {{\it Swift-XRT}}, and $^\ast$ for \Suzaku{}.
(2): Ra in $deg$.
(3): Dec in $deg$.
(4): redshift.
(5): type. References for Sy2s being identified as Compton-thin are ins the superscript brackets, while references for a few activity type changing are in the subscript brackets.
(6): origin \Oline{IV} sample. 1: \citet{Tommasin10}, 2: \citet{Diamond09}, 3: \citet{Weaver10}, 4: \citet{Dasyra11}, 5: \citet{Pereira10}.
(7): \Oline{IV} luminosity in $erg/s$.
(8): absorption corrected 2-10 keV luminosity in $erg/s$. Numbers in the brackets are reference indices.
(9): 14-195 keV luminosity in $erg/s$.
}
\tablerefs{
(1)\citealp{Gallo13_Mrk335};
(2)\citealp{Noguchi09};
(3)\citealp{Tan12_00521};
(4)\citealp{Marchese12_NGC454};
(5)\citealp{Lohfink12_Fariall9};
(6)\citealp{Shu08};
(7)\citealp{hxx11};
(8)\citealp{DeRosa12};
(9)\citealp{Rodriguez08_NGC973};
(10)\citealp{Brenneman09_NGC1052};
(11)\citealp{Nemmen06_NGC1097};
(12)\citealp{Trippe11};
(13)\citealp{Trippe10};
(14)\citealp{Risaliti09_NGC1365};
(15)\citealp{Taiki13_NGC1566_NGC4941};
(16)\citealp{LaMassa11};
(17)\citealp{Longinotti09_H0557-385};
(18)\citealp{Gallo11_Mrk79};
(19)\citealp{LaMassa09};
(20)\citealp{Cappi06};
(21)\citealp{Akylas09};
(22)\citealp{Pappa02_Mrk609};
(23)\citealp{Gallo05_Mrk705};
(24)\citealp{Shu10_NGC2992};
(25)\citealp{Braito07_MCG-5-23-16};
(26)\citealp{Markowitz09_NGC3227};
(27)\citealp{Turner05_NGC3516};
(28)\citealp{Giustini11_Mrk1298};
(29)\citealp{Krongold03_NGC3783};
(30)\citealp{Steenbrugge09_NGC4051};
(31)\citealp{Lubinski10_NGC4151};
(32)\citealp{Turner07_Mrk766};
(33)\citealp{Reynolds09_NGC4258};
(34)\citealp{Beckmann04_NGC4388};
(35)\citealp{Panessa06};
(36)\citealp{Nardini11_NGC4395};
(37)\citealp{Kraft11_NGC4472};
(38)\citealp{Braito13_NGC4507};
(39)\citealp{Li11_NGC4594};
(40)\citealp{Jimenez08_ESO323-G077};
(41)\citealp{Younes11};
(42)\citealp{Miniutti07_IRAS13197};
(43)\citealp{Miller08_MCG-6-30-15};
(44)\citealp{Balestra05_Mrk273};
(45)\citealp{Steenbrugge05_IC4329a};
(46)\citealp{Bianchi08_Mrk463_biAGN};
(47)\citealp{Guainazzi10_NGC5506_broad};
(48)\citealp{Vasudevan13};
(49)\citealp{Shu07};
(50)\citealp{Piconcelli11};
(51)\citealp{Mocz11_IRAS18325};
(52)\citealp{Winter10_NGC6860};
(53)\citealp{Severgnini12};
}
\end{deluxetable*}

\clearpage
\end{CJK*}
\bibliography{ms}
\end{document}